\RequirePackage{fix-cm}
\documentclass[11pt]{article}
\newcounter{ourcount}
\usepackage{amsmath,amsfonts,amssymb,graphicx,epsfig,pdflscape,multirow,theorem,gensymb}
\usepackage[T1]{fontenc}
\usepackage{lscape,arydshln,float}
\usepackage[matrix,arrow,curve]{xy} 
\numberwithin{equation}{section}
\usepackage[nosort]{cite}
\usepackage[usenames]{color}
\usepackage{pstricks}
\usepackage{pst-plot}
\usepackage[backref=false]{hyperref}
\usepackage[capitalise,noabbrev]{cleveref} 
\usepackage{float}
\usepackage[11pt]{moresize}
\hypersetup{
colorlinks=true,
citecolor=red,
linkcolor=darkblue,
urlcolor=darkblue
}
\definecolor{darkblue}{rgb}{0,0,.8}
\definecolor{red}{rgb}{1,0,0}

\theorembodyfont{\itshape} 
\theoremheaderfont{\scshape}
\theoremstyle{plain}

\numberwithin{equation}{section}

\crefname{Conjecture}{Conjecture}{Conjectures}

\newcommand{\nc}{\newcommand}

\def\arxiv#1#2{\href{http://arxiv.org/abs/#1/#2}{\textsf{arXiv:#1/#2}}}
\nc{\ir}{\mathrm{i}}
\nc{\dd}{\mathrm{d}}   
\nc{\eE}{\mathsf{e}}
\nc{\bib}{\bibitem}
\nc{\be}{\begin{equation}}
\nc{\ee}{\end{equation}}
\nc{\bea}{\begin{eqnarray}}
\nc{\eea}{\end{eqnarray}}
\nc{\chit}{\raisebox{0.25ex}{$\chi$}}
\nc{\dtl}{\mathsf{dTL}}
\nc{\pdtl}{\mathsf{pdTL}}
\nc{\Dbh}{\mbox{\boldmath $\widehat D$}}
\nc{\Dh}{\mbox{$\hat D$}}
\nc{\Dbb}{\mbox{\boldmath $\bar D$}}
\nc{\Dbm}{\mbox{\boldmath $\mathcal D$}}
\nc{\Dbt}{\mbox{\boldmath $\tilde{D}$}}
\nc{\Tbt}{\mbox{\boldmath $\tilde{T}$}}
\nc{\Tbh}{\mbox{\boldmath $\widehat{T}$}}

\nc{\setS}{\mathcal S}

\nc{\db}{\mbox{\boldmath $d$}}
\nc{\Ab}{\mbox{\boldmath $A$}}
\nc{\Bb}{\mbox{\boldmath $B$}}
\nc{\Cb}{\mbox{\boldmath $C$}}
\nc{\Db}{\mbox{\boldmath $D$}}
\nc{\eb}{\mbox{\boldmath $e$}}
\nc{\Fb}{\mbox{\boldmath $F$}}
\nc{\Fbt}{\mbox{\boldmath $\tilde{F}$}}
\nc{\fb}{\mbox{\boldmath $f$}}
\nc{\fbt}{\mbox{\boldmath $\tilde{f}$}}
\nc{\Gb}{\mbox{\boldmath $G$}}
\nc{\Hb}{\mbox{\boldmath $H$}}
\nc{\Ib}{\mbox{\boldmath $I$}}
\nc{\Jb}{\mbox{\boldmath $J$}}
\nc{\Kb}{\mbox{\boldmath $K$}}
\nc{\Lb}{\mbox{\boldmath $L$}}
\nc{\Mb}{\mbox{\boldmath $M$}}
\nc{\Pb}{\mbox{\boldmath $P$}}
\nc{\Qb}{\mbox{\boldmath $Q$}}
\nc{\Rb}{\mbox{\boldmath $R$}}
\nc{\Tbb}{\mbox{\boldmath $\bar T$}}
\nc{\Tbm}{\mbox{\boldmath $\mathcal T$}}
\nc{\tb}{\mbox{\boldmath $t$}}
\nc{\Ub}{\mbox{\boldmath $U$}}
\nc{\Vb}{\mbox{\boldmath $V$}}
\nc{\Wb}{\mbox{\boldmath $W$}}
\nc{\xb}{\mbox{\boldmath $x$}}
\nc{\yb}{\mbox{\boldmath $y$}}
\nc{\Zb}{\mbox{\boldmath $Z$}}
\nc{\Lambdab}{\boldsymbol{\Lambda}}
\nc{\T}{\mbox{$\mathbf T$}}
\nc{\Ta}{\mbox{\boldmath $T$}^a}
\nc{\Tb}{\mbox{\boldmath $T$}^b}
\nc{\Tk}{\mbox{\boldmath $T$}^\kappa}
\def\NN#1{\mbox{\boldmath $\mathbf N$}_{#1}}
\def\Nb#1{\mbox{\boldmath $\overline{\mathbf N}$}_{#1}}
\def\nn#1{\mbox{\boldmath $\mathbf n$}_{#1}}
\def\nb#1{\mbox{\boldmath $\bar{\mathbf n}$}_{#1}}

\def\TT{\mbox{\bf T}}

\nc{\even}{\textrm{ even}}
\nc{\odd}{\textrm{ odd}}

\nc{\Atwotwo}{\mbox{$A_2^{\textrm{\fontsize{7pt}{7pt}\selectfont $(2)$}}$}}
\nc{\Aoneone}{\mbox{$A_1^{\textrm{\fontsize{7pt}{7pt}\selectfont $(1)$}}$}}
\nc{\amf}{\mbox{$\mathfrak a$}}
\nc{\bmf}{\mbox{$\mathfrak b$}}
\nc{\cmf}{\mbox{$\mathfrak c$}}
\nc{\dmf}{\mbox{$\mathfrak d$}}
\nc{\fmf}{\mbox{$\mathfrak f$}}
\nc{\gmf}{\mbox{$\mathfrak g$}}
\nc{\Amf}{\mbox{$\mathfrak A$}}
\nc{\Bmf}{\mbox{$\mathfrak B$}}
\nc{\Cmf}{\mbox{$\mathfrak C$}}
\nc{\Dmf}{\mbox{$\mathfrak D$}}
\nc{\Fmf}{\mbox{$\mathfrak F$}}
\nc{\asf}{\mbox{$\mathsf a$}}
\nc{\bsf}{\mbox{$\mathsf b$}}
\nc{\csf}{\mbox{$\mathsf c$}}
\nc{\dsf}{\mbox{$\mathsf d$}}
\nc{\fsf}{\mbox{$\mathsf f$}}
\nc{\gsf}{\mbox{$\mathsf g$}}
\nc{\Asf}{\mbox{$\mathsf A$}}
\nc{\Bsf}{\mbox{$\mathsf B$}}
\nc{\Csf}{\mbox{$\mathsf C$}}
\nc{\Dsf}{\mbox{$\mathsf D$}}

\nc{\repV}{\mathsf{V}}
\nc{\repW}{\mathsf{W}}
\newrgbcolor{darkgreen}{0., 0.733333, 0.0621042}
\definecolor{lightblue}{rgb}{.7,.7,1}
\definecolor{lightestblue}{rgb}{.95,.95,1}
\definecolor{lightlightblue}{rgb}{.85,.85,1}
\definecolor{midblue}{rgb}{.7,.7,1}
\definecolor{lightyellow}{rgb}{1.,.929,.514}
\definecolor{lightorange}{rgb}{.996,.847,.694}
\definecolor{lightpurple}{rgb}{.999,.7,.999}

\nc{\elegant}{1.5pt}
\nc{\moyen}{1.0pt}
\nc{\mince}{0.5pt}

\def\mypmatrix#1{\begin{pmatrix}#1\end{pmatrix}}
\def\smat#1{\mbox{\scriptsize{\mbox{$\mypmatrix{#1}$}}}}
\def\ssmat#1{\mbox{\setlength\arraycolsep{2.5pt}\renewcommand*{\arraystretch}{.95} \scriptsize{\mbox{$\mypmatrix{#1}$}}}}
\def\sssmat#1{\mbox{\setlength\arraycolsep{1.pt}\renewcommand*{\arraystretch}{.95} \scriptsize{\mbox{\tiny$\mypmatrix{#1}$}}}}

\def\vvdots{\mathinner{\mkern1mu\raise1pt\vbox{\kern7pt\hbox{.}}\mkern2mu
  \raise4pt\hbox{.}\mkern2mu\raise7pt\hbox{.}\mkern1mu}}

\def\facegrid#1#2{
\psframe[fillstyle=solid,fillcolor=lightlightblue,linewidth=0pt]#1#2
\psgrid[gridlabels=0pt,subgriddiv=1]#1#2}
\def\facegridy#1#2{
\psframe[fillstyle=solid,fillcolor=lightyellow,linewidth=0pt]#1#2
\psgrid[gridlabels=0pt,subgriddiv=1]#1#2}
\def\facegrido#1#2{
\psframe[fillstyle=solid,fillcolor=lightorange,linewidth=0pt]#1#2
\psgrid[gridlabels=0pt,subgriddiv=1]#1#2}
\def\facegridp#1#2{
\psframe[fillstyle=solid,fillcolor=lightpurple,linewidth=0pt]#1#2
\psgrid[gridlabels=0pt,subgriddiv=1]#1#2}

\def\facegrid#1#2{
\psframe[fillstyle=solid,fillcolor=lightlightblue,linewidth=0pt]#1#2
\psgrid[gridlabels=0pt,subgriddiv=1]#1#2}

\renewcommand{\le}{\leqslant}

\newcommand{\face}[5]{
\psset{unit=0.8cm}
\begin{pspicture}[shift=-.40](0,0)(1,1)
\facegrid{(0,0)}{(1,1)}
\psarc[linewidth=0.025]{-}(0,0){0.16}{0}{90}
\rput(0.,-.1){\spos{tr}{#1}}
\rput(1.,-.1){\spos{tl}{#2}}
\rput(1.,1.1){\spos{bl}{#3}}
\rput(0.,1.1){\spos{br}{#4}}
\rput(.5,.5){$#5$}
\end{pspicture}}
\def\yfacegrid#1#2{
\psframe[fillstyle=solid,fillcolor=lightyellow,linewidth=0pt]#1#2
\psgrid[gridlabels=0pt,subgriddiv=1]#1#2}
\newcommand{\yface}[4]{
\psset{unit=0.6cm}
\begin{pspicture}[shift=-.60](-.3,-.3)(1.3,1.3)
\yfacegrid{(0,0)}{(1,1)}
\psarc[linewidth=0.025]{-}(0,0){0.16}{0}{90}
\rput(0.,-.1){\spos{tr}{#1}}
\rput(1.,-.1){\spos{tl}{#2}}
\rput(1.,1.1){\spos{bl}{#3}}
\rput(0.,1.1){\spos{br}{#4}}
\end{pspicture}}
\newcommand{\braid}{
\psset{unit=0.8cm}
\begin{pspicture}[shift=-.40](0,0)(1,1)
\facegrid{(0,0)}{(1,1)}
\psline[linewidth=1.5pt,linecolor=blue](-.2,.5)(1.2,.5)
\psline[linewidth=1.5pt,linecolor=blue](.5,-.2)(.5,.4)
\psline[linewidth=1.5pt,linecolor=blue](.5,.6)(.5,1.2)
\end{pspicture}}
\newcommand{\invbraid}{
\psset{unit=0.8cm}
\begin{pspicture}[shift=-.40](0,0)(1,1)
\facegrid{(0,0)}{(1,1)}
\psline[linewidth=1.5pt,linecolor=blue](.5,-.2)(.5,1.2)
\psline[linewidth=1.5pt,linecolor=blue](-.2,.5)(.4,.5)
\psline[linewidth=1.5pt,linecolor=blue](.6,.5)(1.2,.5)
\end{pspicture}}

\nc{\proof}{{\scshape Proof.\ }} 				
\nc{\eproof}{{\hfill \rule{0.5em}{0.5em}\medskip}}		
\def\Exp{{\rm Exp}}
\newcommand{\p}[2]{\makebox(0,0)[#1]{$#2$}}
\newcommand{\pp}[2]{\makebox(0,0)[#1]{$\ss#2$}}
\newcommand{\ade}{\mbox{$A$-$D$-$E$ }}
\renewcommand{\ss}{\scriptstyle}

\def\Wt#1#2#3#4#5{W\Big(\!\begin{array}{cc}#4&#3\\#1&#2\end{array}\!\Big|\,#5\Big)}
\def\Wtt#1#2#3#4#5{#1\Big(\!\begin{array}{cc}#5&#4\\#2&#3\end{array}\Big)}
\nc{\spos}[2]{\makebox(0,0)[#1]{$\sm{#2}$}}
\nc{\sm}[1]{{\scriptstyle #1}}
\newcommand{\rotface}[5]{
\psset{unit=0.8cm}
\begin{pspicture}[shift=-.40](0,0)(1,1)
\facegrid{(0,0)}{(1,1)}
\psarc[linewidth=0.025]{-}(0,1){0.16}{270}{360}
\rput(0.,-.1){\spos{tr}{#1}}
\rput(1.,-.1){\spos{tl}{#2}}
\rput(1.,1.1){\spos{bl}{#3}}
\rput(0.,1.1){\spos{br}{#4}}
\rput(.5,.5){$#5$}
\end{pspicture}}

\begin{document}

\topmargin -5mm
\oddsidemargin 5mm
\vspace*{-2cm}
%\begin{flushright}
%APCTP Pre2014-010\\
%\jobname .pdf\ \today
%\end{flushright}

\makeatletter 
\newcommand\Larger{\@setfontsize\semiHuge{20.00}{23.78}}
\makeatother 

\setcounter{page}{1}
\mbox{}\vspace{-.2cm}\mbox{}
\begin{center}

{\Larger \bf \mbox{
%{\color{red}modular invariant} 
Ocneanu Algebra of Seams:\ \ }
\\[0.18cm] 
\mbox{Critical Unitary 
%$(A_{10},E_6)$ 
$E_6$ RSOS Lattice Model}
%\\[.3cm]
%\mbox{periodic and antiperiodic boundary conditions}
}

\end{center}

\vspace{1cm}
\begin{center}
{\vspace{-5mm}\Large Paul A.~Pearce$^{a,b}$, J\o{}rgen Rasmussen$^b$}
\\[.5cm]
{\em { }$^a$School of Mathematics and Statistics, University of Melbourne\\
Parkville, Victoria 3010, Australia}
\\[.2cm]
{\em { }$^b$School of Mathematics and Physics, University of Queensland}\\
{\em St Lucia, Brisbane, Queensland 4072, Australia}
\\[.2cm] 
\qquad
{\tt papearce\,@unimelb.edu.au}
\qquad
{\tt j\!.rasmussen\,@uq.edu.au}
\end{center}

\vspace{4mm}
\centerline{{\bf{Abstract}}}
\vskip.3cm
\noindent 
We consider the $A$ series and exceptional $E_6$ Restricted Solid-On-Solid lattice models as prototypical examples of the critical Yang-Baxter integrable two-dimensional $A$-$D$-$E$ lattice models. We focus on type I theories which are characterized by the existence of an extended chiral symmetry in the continuum scaling limit. 
Starting with the commuting family of column transfer matrices on the torus, we build matrix representations of the Ocneanu graph fusion algebra as integrable seams for arbitrary finite-size lattices. 
In the $A$ cases, the Ocneanu algebra coincides with the Verlinde algebra.
In the other cases, the seam algebra contains the fused adjacency and graph fusion algebras as subalgebras. Our matrix representation of the Ocneanu algebra encapsulates the quantum symmetry of the commuting family of transfer matrices.
In the continuum scaling limit, the integrable seams realize the topological defects of the associated conformal field theory and the known toric matrices encode the twisted conformal partition functions of Petkova and Zuber. 
\vspace{-.4cm}
%\newpage
\tableofcontents

\newpage
\hyphenpenalty=30000

\setcounter{footnote}{0}

%%%%%%%%%%%%%%%%%%%%%%%%%%%%
%
\section{Introduction}
%
%%%%%%%%%%%%%%%%%%%%%%%%%%%%

In this short article, we are concerned with two-dimensional critical Restricted Solid-On-Solid (RSOS) lattice models built on the Dynkin diagrams of simply-laced $sl(2)$ \ade Lie algebras as in Figure~\ref{fig:Graphs}. These models are Yang-Baxter integrable~\cite{BaxBook82}. The first such models were  built on the $A$-type Dynkin diagrams and solved off-criticality in 1984 by Andrews, Baxter and Forrester~\cite{ABF84}. The full family of critical \ade lattice models was subsequently introduced and studied by Pasquier~\cite{Pasquier87a,Pasquier87b,Pasquier87c}. In the continuum scaling limit, the thermodynamic behaviour of these models is described by Conformal Field Theory (CFT)~\cite{FMS97}. The $A$ series is associated with the conformal unitary minimal models \mbox{${\cal M}(g\!-\!1,g)$} of Belavin, Polyakov and Zamolodchikov~\cite{BPZ84}. 
The unitary \ade CFTs are in fact $(A,G)$ coset models~\mbox{\!\!\cite{GKO85,FMS97}}
\be
(A_{g-2},G)=\begin{cases}
(A_{g-2},A_{g-1}),&g=4,5,6,\ldots\\
(A_{g-2},D_{(g+2)/2}),&g=6,8,10,\ldots\\
(A_{10},E_6),\ \ \  &g=12\\
(A_{16},E_7),\ \ \ &g=18\\
(A_{28},E_8),\ \ \ &g=30
\end{cases}\qquad\ \ \   c=1-\frac{6}{g(g-1)}
\ee
where $g$ is the Coxeter number of $G$ and $c$ is the central charge. These theories are rational~\cite{MooreSeiberg}, admitting a finite number of irreducible representations of the Virasoro algebra which close under fusion. 

\begin{figure}[htb]
\begin{center}
    \setlength{\unitlength}{.5cm}
\begin{equation*}
\mbox{}\hspace{1.4in}\mbox{}
\begin{array}{ccccc}
   \text{\text{Graph $G$\qquad}}& \mbox{}\quad\qquad\text{$g$}\quad\qquad\mbox{}&\text{$\Exp(G)$}&\text{Type/$H$}&\Gamma\\[10pt]
   \begin{picture}(6,1)
   	\rput(-1.8,1.1){Lie Algebra}
	\psline[linewidth=.5pt](-3.6,.7)(13.3,.7)
	\psline[linewidth=.5pt](-3.6,-7.3)(13.3,-7.3)
        \put(-3.5,0){\p{}{A_{L}}}
     % The line
        \put(0,0){\line(1,0){4}}
     % The dots
        \multiput(1,0)(1,0){2}{\pp{}{\bullet}}
        \put(4,0){\pp{}{\bullet}}
     % The numbers
        \put(0,0){\pp{}{\bullet}}
      %  \put(1,0){\pp{}{\square}}
       \put(0,.3){\pp{b}{1}}
        \put(1,.3){\pp{b}{2}}
        \put(2,.3){\pp{b}{3}}
        \put(3,.3){\pp{b}{\cdots}}
        \put(4,.3){\pp{b}{L}}
    \end{picture}  & L+1 & 1, 2, \ldots, L &\mbox{I}&{\Bbb Z_2}\\
     \begin{picture}(6,2)
        \put(-3.5,0){\p{}{D_L\,\mbox{($L$ even)}}}
     % The lines
        \put(0,0){\line(1,0){3.5}}
        \put(3.5,0){\line(1,1){1}}
        \put(3.5,0){\line(1,-1){1}}
     % The dots
        \multiput(1,0)(1,0){2}{\pp{}{\bullet}}
        \put(3.5,0){\pp{}{\bullet}}
        \put(4.5,1){\pp{}{\bullet}}
        \put(4.5,-1){\pp{}{\bullet}}
    % The numbers
        \put(-.2,-.04){\pp{l}{\bullet}}
      %  \put(.775,-.00){\pp{l}{\square}}
        \put(0,.3){\pp{b}{1}}
        \put(1,.3){\pp{b}{2}}
        \put(2,.3){\pp{b}{3}}
        \put(2.75,.3){\pp{b}{\cdots}}
        \put(3.5,.3){\pp{b}{L\!-\!2}}
        \put(4.5,1){\pp{l}{~L\!-\!1}}
        \put(4.5,-.7){\pp{l}{~L}}
       % \put(4.35,-1.05){\pp{l}{*_{\text{od}}}}
        %\put(3.25,-.05){\pp{l}{\square_{\text{od}}}}
    \end{picture}    & 2L -2 & 1, 3, \ldots, 2L-3, L-1 &\mbox{I}&{\Bbb Z_2/\Bbb S_3} \\[10pt]
 \begin{picture}(6,2)
        \put(-3.5,0){\p{}{D_{L}\,\mbox{($L$ odd)}}}
     % The lines
        \put(0,0){\line(1,0){3.5}}
        \put(3.5,0){\line(1,1){1}}
        \put(3.5,0){\line(1,-1){1}}
     % The dots
        \multiput(0,0)(1,0){3}{\pp{}{\bullet}}
        \put(3.5,0){\pp{}{\bullet}}
        \put(4.5,1){\pp{}{\bullet}}
        % \put(4.5,-1){\pp{}{\bullet}}
    % The numbers
        %\put(-.2,-.04){\pp{l}{*_{\text{ev}}}}
       % \put(.75,-.04){\pp{l}{\square_{\text{ev}}}}
        \put(0,.3){\pp{b}{1}}
        \put(1,.3){\pp{b}{2}}
        \put(2,.3){\pp{b}{3}}
        \put(2.75,.3){\pp{b}{\cdots}}
        \put(3.5,.3){\pp{b}{L\!-\!2}}
        \put(4.5,1){\pp{l}{~L\!-\!1}}
        \put(4.5,-.7){\pp{l}{~L}}
        \put(4.35,-1.05){\pp{l}{\bullet}}
     %   \put(3.275,-.00){\pp{l}{\square}}
    \end{picture}    & 2L-2 & 1, 3, \ldots, 2L-3, L-1 &\mbox{II}/A_{2L-3} &{\Bbb Z_2}\\
   \begin{picture}(6,2.5)
        \put(-3.5,0){\p{}{E_{6}}}
     % The line
        \put(0,0){\line(1,0){4}}
        \put(2,0){\line(0,1){1}}
     % The dots
        \multiput(1,0)(1,0){4}{\pp{}{\bullet}}
        \put(2,1){\pp{}{\bullet}}
     % The numbers
        \put(0,0){\pp{}{\bullet}}
     %   \put(1,0){\pp{}{\square}}
        \put(0,.3){\pp{b}{1}}
        \put(1,.3){\pp{b}{2}}
        \put(2,-.3){\pp{t}{3}}
        \put(3,.3){\pp{b}{4}}
        \put(4,.3){\pp{b}{5}}
        \put(2,1.3){\pp{b}{6}}
     \end{picture}  & 12 & 1, 4, 5, 7, 8, 11 &\mbox{I}&{\Bbb Z_2} \\
   \begin{picture}(6,2.5)
        \put(-3.5,0){\p{}{E_{7}}}
     % The line
        \put(0,0){\line(1,0){5}}
        \put(3,0){\line(0,1){1}}
     % The dots
        \multiput(1,0)(1,0){5}{\pp{}{\bullet}}
         \put(3,1){\pp{}{\bullet}}
     % The numbers
        \put(0,.3){\pp{b}{1}}
        \put(1,.3){\pp{b}{2}}
        \put(2,.3){\pp{b}{3}}
        \put(3,-.3){\pp{t}{4}}
        \put(4,.3){\pp{b}{5}}
        \put(5,.3){\pp{b}{6}}
        \put(3,1.3){\pp{b}{7}}
        \put(0,0){\pp{}{\bullet}}
      %  \put(1,0){\pp{}{\square}}
     \end{picture}  & 18 & 1, 5, 7, 9, 11, 13, 17  &\mbox{II}/D_{10}&1\\
   \begin{picture}(6,2.5)
        \put(-3.5,0){\p{}{E_{8}}}
     % The line
        \put(0,0){\line(1,0){6}}
        \put(4,0){\line(0,1){1}}
     % The dots
        \multiput(1,0)(1,0){6}{\pp{}{\bullet}}
        \put(4,1){\pp{}{\bullet}}
     % The numbers
        \put(0,0){\pp{}{\bullet}}
      %  \put(1,0){\pp{}{\square}}
        \put(0,.3){\pp{b}{1}}
        \put(1,.3){\pp{b}{2}}
        \put(2,.3){\pp{b}{3}}
        \put(3,.3){\pp{b}{4}}
        \put(4,-.3){\pp{t}{5}}
        \put(5,.3){\pp{b}{6}}
        \put(6,.3){\pp{b}{7}}
        \put(4,1.3){\pp{b}{8}}
     \end{picture}  & 30 & 1, 7, 11, 13, 17, 19, 23, 29&\mbox{I}&1\\
\end{array}
\end{equation*}
\end{center}
\caption{Dynkin diagrams of the classical simply-laced $sl(2)$ \ade 
Lie algebras. The nodes associated with the identity and the fundamental are
labelled 1 and 2 respectively. The fundamental is the unique neighbour of the identity. Also shown are the Coxeter numbers $g$, exponents $\Exp(G)$, the type I or II and the so-called parent graphs $H\ne G$. The diagram automorphism group $\Gamma$ is generated by a single $\mathbb{Z}_2$ automorphism $\sigma$. The $D_4$ graph is an exception having the noncommutative automorphism group $\Bbb S_3$. The eigenvalues of $G$ are $2\cos \tfrac{s\pi}{g}$ with $s\in \Exp(G)$. 
By abuse of notation, we use $G$ to denote the graph, the set of its vertices with cardinality $|G|$ and its adjacency matrix so that $2I-G$ is the Cartan matrix. The meaning of $G$ should be clear from context.}
\label{fig:Graphs}
\end{figure}

Much effort within CFT has been focussed on the study of boundary conditions. On the cylinder, the boundary conditions of unitary minimal CFTs and their conformal cylinder partition functions are understood~\cite{BPPZ1998,BPPZ} in terms of nonnegative integer matrix representations (nimreps) of the Verlinde~\cite{Verlinde88} and graph fusion~\cite{Pasquier87b} algebras. The irreducible representations are conjugate to conformal boundary conditions~\cite{BP2001} labelled by the coset graph nodes $(r,a)\in (A,G)$. Much of this structure is common \cite{PasSal90,BPZ1998} between the finite lattice models and the associated CFT. 
On the torus with periodic boundary conditions in both directions, the  modular invariant partition functions~\cite{CardyModInv86} with $c<1$ are exhausted by the \ade classification of Cappelli, Itzykson and Zuber~\cite{CIZ87}. Incorporating the left and right chiral copies of the Virasoro algebra, the conformal partition functions are given as sesquilinear forms in Virasoro characters. 
Twisted boundary conditions are imposed by inserting topological defect seams into the bulk. 
The resulting twisted partition functions on the torus are understood in terms of the Ocneanu~\cite{Ocneanu} fusion algebra which describes the quantum symmetry and how the left and right chiral halves of the theory are glued together. Indeed, working within CFT, Petkova and Zuber~\cite{PetkovaZuber2001} have obtained explicit expressions for 
the toric matrices and the associated twisted conformal partition functions. The algebraic structure behind these results has been elucidated by Coquereaux and collaborators~\cite{Coquereaux2000,CoqSchieber2002,CoqHuerta2003,CoqTrinchero} who, following Ocneanu~\cite{Ocneanu}, identified the Ocneanu algebra with the quotient algebra $G\otimes_T G:=(G\otimes G)\slash T$ where $G$ denotes the graph algebra of type $A,D,E$ and where $T$ is the ambichiral subalgebra of $G$ specified in \cite{PetkovaZuber2001}. In particular, the authors of \cite{Coquereaux2000,CoqSchieber2002} give a detailed account of the prototypical type~I example of $E_6$.

In parallel to the developments coming from CFT, twisted boundary conditions on the torus have also been studied from a purely lattice perspective~\cite{ChuiEtAl2001,ChuiEtAlOdyssey2001,ChuiEtAl2003} by inserting integrable seam segments into the commuting {\em row} transfer matrices. These segments give rise to commuting integrable {\em vertical} seams on the square lattice which, in the continuum scaling limit, become the topological defects of the associated CFT. In this picture, the fusion of the integrable seams is simply implemented by matrix products of the integrable seams. 
This picture was implicitly confirmed in \cite{ChuiEtAl2003} by numerical studies of the conformal spectra of the commuting row transfer matrices.

The aim of this article is to develop further the lattice approach to fusion of integrable seams on the lattice. As in \cite{BelleteteEtAl2023,TavaresEtAl2024}, we turn our attention to the ``cross-channel'', but here we use the commuting family of column transfer matrices to build the various types of integrable seams on the torus. 
Modifying the notation of \cite{PetkovaZuber2001,ChuiEtAl2003}, the composite integrable seams are labelled by $x=(r,\kappa,a,b)$. The associated {\em column} transfer matrices are constructed by matrix products of elementary seam transfer matrices
\be
{\mathbf T}_x(u)={\mathbf T}_r(u,\xi)\,{\boldsymbol\sigma}^\kappa\,{\widehat{\mathbf N}_a}\,{\overline{\widehat{\mathbf N}}_b},\qquad x=(r,\kappa,a,b),\qquad r\in A_{g-2},\ \ \kappa=0,1,\ \ a,b\in G
\ee
as shown in Figure~\ref{TMs}. Here ${\mathbf T}_r(u,\xi)$ is the column transfer matrix for an $r$-type seam, $\xi$ is a column inhomogeneity, ${\widehat{\mathbf N}_a},\,{\overline{\widehat{\mathbf N}}_b}$ are graph fusion seams labelled by $a,b\in G$ (as defined in Section~\ref{sec:Ocneanu}) and ${\boldsymbol\sigma}^\kappa={\mathbf I}$ or ${\boldsymbol\sigma}$ where the $\sigma$-parity 
$\kappa=0,1$ labels respectively the identity ${\mathbf I}$ and the $\mathbb{Z}_2$ graph automorphism seam ${\boldsymbol\sigma}$. 
The seam labels $\kappa$ do not need to be explicitly added in $x$ except for the $D_L$ ($L$ even) series of lattice models.

As pointed out in \cite{ChuiEtAl2003}, due to the coset factorization structure~\cite{GKO1985}, the $r$-type seams (with {$r\in A_{g-2}$}) play a somewhat trivial role. We will therefore usually set $r=1$ (no $r$-type seam) and focus on the so-called Wess-Zumino-Witten (WZW) factor associated with the graph $G$. For $x=(a,b)$, we assert that the integrable seams $\widehat{\mathbf P}_{a,b}={\widehat{\mathbf N}_a}\,{\overline{\widehat{\mathbf N}}_b}$ yield matrix representations of the Ocneanu fusion algebra for arbitrary system sizes. In particular, for the critical \ade lattice models, the Ocneanu algebra encodes the quantum symmetries possessed by the commuting families of column transfer matrices on a finite lattice.
Following \cite{Coquereaux2000}, we focus here on the $A$ series and the prototypical type~I example of $E_6$. Our full findings for the  $D$ and other $E$ critical RSOS lattice models will appear elsewhere.

\begin{figure}[htb]
\begin{equation*}
\psset{unit=.9cm}
%\fbox{
\begin{pspicture}[shift=-3.9](-2,-1)(8,7.75)
\multirput(0,0)(0,1){8}{\psline[linewidth=1.pt,linestyle=dashed,dash =2pt 1pt](-2.5,-.5)(8.5,-.5)}
\multirput(0,0)(1,0){10}{\psline[linewidth=1.pt,linestyle=dashed,dash =2pt 1pt](-1.5,-1.5)(-1.5,7.5)}
\facegrid{(-2,-1)}{(8,7)}
\multirput(0,0)(0,1){8}{\multirput(0,0)(1,0){3}{\rput(-1.5,-.5){$u$}}}
\multirput(4,0)(0,1){8}{\multirput(0,0)(1,0){6}{\rput(-1.5,-.5){$u$}}}
\multirput(3,0)(0,1){8}{\multirput(0,0)(1,0){1}{\rput(-1.5,-.5){$-i\infty$}}}
\facegridy{(5,-1)}{(4,7)}
\facegrido{(6,-1)}{(5,7)}
\facegridp{(6,-1)}{(7,7)}
%\facegridb{(7,-1)}{(8,7)}
\psframe[fillstyle=none,linecolor=red,linewidth=1.5pt](-2,6)(8,7)
\psframe[fillstyle=none,linewidth=1.pt](4,-1)(5,7)
\psframe[fillstyle=none,linewidth=1.pt](5,-1)(6,7)
\psframe[fillstyle=none,linewidth=1.pt](6,-1)(7,7)
%\psframe[fillstyle=none,linewidth=1.pt](7,-1)(8,7)
\psframe[fillstyle=none,linewidth=1.5pt](-1,-1)(0,7)
\multirput(0,0)(0,1){8}{\multirput(0,0)(1,0){10}{\psarc[linewidth=0.025]{-}(-2,-1){0.16}{0}{90}}}
\rput(-3.2,6.5){$\T_h^x(u)$}
\rput(-.5,-1.8){$\T(u)$}
\rput(1.5,-1.8){$\nn s$}
\rput(4.6,-1.84){$\,\,\boldsymbol\sigma^\kappa$}
\rput(5.55,-1.8){$\widehat{\mathbf N}_a$}
\rput(6.5,-1.77){$\,\overline{\widehat{\mathbf N}}_b$}
\multirput(0,0)(0,1){8}{\rput(4.5,-.5){$W_\kappa$}}
\multirput(1,.0)(0,1){8}{\rput(4.5,-.5){$W_a$}}
\multirput(2,0)(0,1){8}{\rput(4.5,-.5){$W_b$}}
\psline[linewidth=1.75pt](4,-1)(4,7)
\psline[linewidth=1.75pt](7,-1)(7,7)
%}
\end{pspicture}
\end{equation*}\\[0pt]
\caption{An $N\times M$ periodic lattice with $(N,M)=(10,8)$ showing (i) the row transfer matrix $\T_h^x(u)$ with seam segment $x=(\kappa,a,b)$, (ii) the column/seam transfer matrices $\T(u)$, $\nn s$, $\widehat{\mathbf N}_a$ and $\overline{\widehat{\mathbf N}}_b$ as explained in Section~\ref{sec:Ocneanu} and (iii) the ${\mathbb Z}_2$ diagram automorphism seam $\boldsymbol\sigma^\kappa$. The composite seam is the matrix product $\TT_x=\boldsymbol\sigma^\kappa\,\widehat{\mathbf N}_a\,\overline{\widehat{\mathbf N}}_b$. The seam $\TT_x$ has the same twisted partition function as the seam
$\widehat{\mathbf N}_a\,\overline{\widehat{\mathbf N}}_b\,\boldsymbol\sigma^\kappa$. We assume $r=1$ so there is no $r$-type seam shown. The labels $W_\kappa, W_a$ and $W_b$ indicate that special face weights are assigned to these faces.}
\label{TMs}
\end{figure}

%%%%%%%%%%%%%%%%%%%%%%%%%%%%
%
\section{Critical RSOS models on the torus}\label{sec:TransferMatrices}
%
%%%%%%%%%%%%%%%%%%%%%%%%%%%%

\def\vec#1{\boldsymbol{#1}}

\subsection{Critical \ade lattice models}

Defining quantum dimensions by $S_a=[a]_{x}=\frac{x^{a}-x^{-a}}{x-x^{-1}}$  with $x=e^{i\lambda}$, the nondegenerate largest eigenvalue of the adjacency matrix $G$ is \mbox{$[2]_{x}=2\cos\lambda$} and the associated (unnormalized) Perron-Frobenius eigenvector $\boldsymbol{\psi}$ is 
\be
G\vec\psi=[2]_x\,\vec\psi,\qquad
\vec\psi=(\psi_a)_{1\le a\le |G|}=\begin{cases}
\bigl( [1]_{x},[2]_x,\ldots,[L]_x\bigr),&G=A_L\\[4pt]
\bigl( [1]_{x}, [2]_x,\ldots,[\ell]_x,\frac{[\ell]_{x}}{[2]_{x}},\frac{[\ell]_{x}}{[2]_{x}}\bigr),&G=D_{\ell+2}\\[4pt]
\bigl( [1]_{x}, [2]_{x}, [3]_{x}, [2]_{x}, [1]_{x},\frac{[3]_{x}}{[2]_{x}}\bigr),&G=E_6\\[4pt]
\bigl( [1]_{x}, [2]_{x}, [3]_{x}, [4]_{x}, \frac{[6]_{x}}{[2]_{x}},\frac{[4]_{x}}{[3]_{x}},\frac{[4]_{x}}{[2]_{x}}\bigr),&G=E_7\\[4pt]
\bigl( [1]_{x}, [2]_{x}, [3]_{x}, [4]_{x},  [5]_{x}, \frac{[7]_{x}}{[2]_{x}},\frac{[5]_{x}}{[3]_{x}},\frac{[5]_{x}}{[2]_{x}}\bigr),&G=E_8
\end{cases}
\ee
The allowed face weights of the critical $A$-$D$-$E$ lattice models are then given by
\begin{equation}
\psset{unit=0.8cm}
\Wt abcdu=\;\face abcdu\;
=\frac{\sin(\lambda-u)}{\sin\lambda}\,\delta_{ac}+
    \frac{\sin u}{\sin\lambda}\,\sqrt{\frac{\psi_{a}\psi_{c}}{\psi_{b}\psi_{d}}}\;\delta_{bd},\quad\  \lambda=\frac{\pi}{g},\quad\  0\le u\le \lambda
           \label{eq:W11}
\end{equation}
where $u$ is the spectral parameter, $\lambda$ is the crossing parameter and the face orientation is marked by an arc in the bottom-left corner. 
It is understood that the face weights vanish if $G_{ab}G_{bc}G_{cd}G_{da}=0$. 
The face weights are invariant under reflections in the two diagonals and satisfy the crossing symmetry
\be
\Wt abcdu=\;\face abcd{\mbox{\small $u$}}\;=\;\sqrt{\frac{\psi_a\psi_c}{\psi_b\psi_d}}\ \ \rotface abcd{\mbox{\small $\lambda\!-\!u$}}\;=\sqrt{\frac{\psi_a\psi_c}{\psi_b\psi_d}}\ \Wt dabc{\lambda\!-\!u}
\ee

\subsection{Transfer matrices, braid limit and Hamiltonians}
The row and column families of commuting transfer matrices $\T_{\!h}(u),\T(u)$ of the critical \ade models on an $N\times M$ periodic lattice are defined by
\psset{unit=0.8cm}
\begin{align}
\T_{\!h}(u)
&= \ \label{Th}
\begin{pspicture}[shift=-0.9](-0.3,-0.5)(5.3,1.2)
\facegrid{(0,0)}{(5,1)}
\psarc[linewidth=0.025]{-}(0,0){0.16}{0}{90}
\psarc[linewidth=0.025]{-}(1,0){0.16}{0}{90}
\psarc[linewidth=0.025]{-}(2,0){0.16}{0}{90}
\psarc[linewidth=0.025]{-}(4,0){0.16}{0}{90}
\psline[linewidth=1.pt,linecolor=black,linestyle=dashed,dash=2pt 1pt]{-}(0,0.5)(-0.3,0.5)
\psline[linewidth=1.pt,linecolor=black,linestyle=dashed,dash=2pt 1pt]{-}(5,0.5)(5.3,0.5)
\rput(0.5,.5){$u$}
\rput(1.5,.5){$u$}
\rput(2.5,0.5){$u$}
\rput(3.5,0.5){$\ldots$}
\rput(4.5,.5){$u$}
\psline{<->}(0,-0.2)(5,-0.2)\rput(2.5,-0.45){$_N$}
\end{pspicture}\\[4pt]
%\psset{unit=0.6cm}
\T(u)=\T_{\!v}(u)
%&=\T_{\!h}(\lambda-u)
&= \ 
\label{Tv}\begin{pspicture}[shift=-0.9](-0.3,-0.5)(5.3,1.2)
\facegrid{(0,0)}{(5,1)}
\multirput(0,0)(1,0){3}{\psarc[linewidth=0.025]{-}(0,1){0.16}{270}{360}}
\psarc[linewidth=0.025]{-}(4,1){0.16}{270}{360}
\psline[linewidth=1.pt,linecolor=black,linestyle=dashed,dash=2pt 1pt]{-}(0,0.5)(-0.3,0.5)
\psline[linewidth=1.pt,linecolor=black,linestyle=dashed,dash=2pt 1pt]{-}(5,0.5)(5.3,0.5)
\rput(0.5,.5){$u$}
\rput(1.5,.5){$u$}
\rput(2.5,0.5){$u$}
\rput(3.5,0.5){$\ldots$}
\rput(4.5,.5){$u$}
\psline{<->}(0,-0.2)(5,-0.2)\rput(2.5,-0.45){$_M$}
\end{pspicture}
\end{align}
where, for convenience, the column transfer matrix is rotated clockwise by 90 degrees.
The equality $\T_{\!v}(u)=\T_{\!h}(\lambda-u)$ holds when $N=M$. 
Since $[\T(u),\T(v)]=\mathbf 0$ and $\T(u)^T=\T(\lambda\!-\!u)$ by crossing symmetry, 
the family of normal matrices $\T(u)$ is simultaneously diagonalizable by a unitary matrix. The corresponding eigenvectors are independent of $u$ and in general complex. 
Other column transfer matrices corresponding to integrable seams are shown in Figure~\ref{TMs}.

\begin{subequations}
The braid limits of the allowed \ade face weights are
\begin{align}\label{braid}
\psset{unit=0.8cm}
\Wtt Babcd&=\ 
\begin{pspicture}[shift=-.40](0,0)(1,1)
\facegrid{(0,0)}{(1,1)}
\psline[linewidth=1.5pt,linecolor=blue](-.2,.5)(1.2,.5)
\psline[linewidth=1.5pt,linecolor=blue](.5,-.2)(.5,.4)
\psline[linewidth=1.5pt,linecolor=blue](.5,.6)(.5,1.2)
\rput(0.,-.1){\spos{tr}{a}}
\rput(1.,-.1){\spos{tl}{b}}
\rput(1.,1.1){\spos{bl}{c}}
\rput(0.,1.1){\spos{br}{d}}
\end{pspicture}
\ =- \lim_{u\to-i\infty}\frac{x^{\frac 12}}{\rho(u)} \,\Wt abcdu
=i\bigg(\!\!-\!x^{-\frac 12}\,\delta_{ac}+x^{\frac 12}\sqrt{\frac{\psi_{a}\psi_{c}}{\psi_{b}\psi_{d}}}\;\delta_{bd}\bigg)\\[6pt]
\label{invbraid}
\Wtt {\overline{B}}abcd&=\ 
\begin{pspicture}[shift=-.40](0,0)(1,1)
\facegrid{(0,0)}{(1,1)}
\psline[linewidth=1.5pt,linecolor=blue](.5,-.2)(.5,1.2)
\psline[linewidth=1.5pt,linecolor=blue](-.2,.5)(.4,.5)
\psline[linewidth=1.5pt,linecolor=blue](.6,.5)(1.2,.5)
\rput(0.,-.1){\spos{tr}{a}}
\rput(1.,-.1){\spos{tl}{b}}
\rput(1.,1.1){\spos{bl}{c}}
\rput(0.,1.1){\spos{br}{d}}
\end{pspicture}
\ =- \lim_{u\to i\infty}\;\,\frac{x^{-\frac 12}}{\rho(u)}\;\, \Wt abcdu
=i\bigg(x^{\frac 12}\,\delta_{ac}\;-\;x^{-\frac 12}\sqrt{\frac{\psi_{a}\psi_{c}}{\psi_{b}\psi_{d}}}\;\delta_{bd}\bigg)
\end{align}
where $x=e^{i\lambda}$ and $\rho(u)=\sin(\lambda\!-\!u)\sin(\lambda\!+\!u)$. These braid limits relate to the left and right chiral halves of the theory and are related to each other  by complex conjugation. Taking the action of the face transfer operators $X_j(u)$ to be diagonal and using the Temperley-Lieb generators $e_j$, these expressions become
\end{subequations}\begin{subequations}
\begin{align}
\psset{unit=0.8cm}
b_j&=\;\;\braid\;\;=- \lim_{u\to-i\infty}\frac{x^{\frac 12}}{\rho(u)} \,X_j(u)=i\big(\!-\!x^{-\frac 12} I+x^{\frac 12} e_j\big)\\[6pt]
\overline{b}_j&=\;\;\invbraid\;\;=- \lim_{u\to i\infty}\frac{x^{-\frac 12}}{\rho(u)}\, X_j(u)=i\big(x^{\frac 12} I-x^{-\frac 12} e_j\big)
=b_j^{-1}\label{TLBraid}
\end{align}
\end{subequations}

The Hamiltonian limit of the horizontal transfer matrix $\T_h(u)$ (\ref{Tv}) gives rise to periodic anyonic chains
\be
{\cal H}=-\sin\lambda\, \frac{d}{du}\log\T_h(u) \Big|_{u=0}=-\sum_{j=1}^N e_j
\ee
The simplest case is the Hamiltonian limit of the tricritical Ising $A_4$ RSOS model~\cite{KP1991} corresponding to the Fibonacci golden chain~\cite{Anyons,HeymannQuella}.

%%%%%%%%%%%%%%%%%%%%%%%%%%%%
%
\section{Ocneanu fusion algebra}\label{sec:FusionStructure}
%
%%%%%%%%%%%%%%%%%%%%%%%%%%%%

Most of the content in this section is known~\cite{Ocneanu,PetkovaZuber2001,Coquereaux2000,CoqSchieber2002,CoqHuerta2003,CoqTrinchero}. We review some material to establish notation and to develop a more economical picture using new smaller extended toric matrices rather than the known larger (unextended) toric matrices. This extended picture holds for type I and also type II cases when the parent graph $H$ is of type I.

\subsection{Fusion matrices and structure constants}
The fused adjacency matrices (intertwiners) $n_i$ and graph fusion matrices $\widehat{N}_a$ are defined by the finitely truncated recursions
\begin{align}
n_1&=I,\quad n_2=G,\quad n_i n_2=n_{i-1}+n_{i+1},\quad n_{g}=0,\qquad i\in A_{g-1}\label{nRecursion}\\[4pt]
\widehat{N}_1&=I,\quad \widehat{N}_2=G,\quad G\,\widehat{N}_a=\sum_{b\in G}G_{ab}\,\widehat{N}_b,\qquad a\in G\label{GRecursion}
\end{align}
For $G=A_{g-1}$, the fused adjacency matrices $n_i$ are in fact the Verlinde~\cite{Verlinde88} fusion matrices $N_i$. The fused adjacency and graph fusion matrices satisfy the commutative associative fusion algebras
\be
n_i\,n_j=\sum_{k=1}^{g-1} N_{ij}{}^k\,n_k,\qquad 
\widehat{N}_a\,\widehat{N}_b=\sum_{c\in G} \widehat{N}_{ab}{}^c\,\widehat{N}_c,\qquad n_i\,\widehat{N}_a=\sum_{b\in G} n_{ia}{}^b\,\widehat{N}_b\label{FusionAlgebras}
\ee
where the structure constants are
\be
N_{ij}{}^k=(N_i)_j{}^k\in{\Bbb N},\qquad \widehat{N}_{ab}{}^c=(\widehat{N}_a)_b{}^c\in\begin{cases} {\Bbb N},&\mbox{type I}\\ {\Bbb Z},&\mbox{type II}\end{cases},\qquad n_{ia}{}^b=(n_i)_a{}^b\in{\Bbb N}
\ee
For type I theories, the $N_i$, $\widehat{N}_a$ and $n_i$ form nimreps (nonnegative integer matrix representations) of these fusion algebras. The type II theories do not admit proper graph fusion algebras.

Denoting the complex unitary matrices of eigenvectors of $N_2$, $\widehat{N}_2$ by $S_i{}^\ell$ (modular matrix) and $\Psi_a{}^\ell$ with $S_1{}^\ell,\Psi_1{}^\ell>0$, the corresponding Verlinde and Verlinde-type formulas are
\be
 N_{ij}{}^k=\sum_{\ell=1}^{g-1}{S_i{}^\ell\, S_j{}^\ell\, \overline{S_k{}^\ell}\over S_1{}^\ell},\quad 
 \widehat{N}_{ab}{}^c=\sum_{\ell\in {\rm Exp}(G)} 
{\Psi_a{}^\ell \Psi_b{}^\ell \,\overline{\Psi_c{}^\ell} \over \Psi_1{}^\ell},\quad
 n_{ia}{}^b=\sum_{\ell\in\Exp(G)} {S_i{}^\ell\,\Psi_a{}^\ell\, \overline{\Psi_b{}^\ell}\over S_1{}^\ell}
\ee
Here bars denote complex conjugation and the sum in the first expression is over $\ell\in{\rm Exp}(A_{g-1})$. 
For structure constants larger than 1, it is necessary to introduce so-called bond variables (fusion graphs with multiple edges). 
Explicit fusion adjacency matrices for $A_4$ and $E_6$ are
\begin{align}
N_1&=\smat{1&0&0&0\\ 0&1&0&0\\ 0&0&1&0\\ 0&0&0&1},\quad 
N_2=\smat{0&1&0&0\\ 1&0&1&0\\ 0&1&0&1\\ 0&0&1&0},\quad 
N_3=\smat{0&0&1&0\\ 0&1&0&1\\ 1&0&1&0\\ 0&1&0&0},\quad 
N_4=\smat{0&0&0&1\\ 0&0&1&0\\ 0&1&0&0\\ 1&0&0&0}\ \;\\[4pt]
\widehat{N}_1&=\smat{1&0&0&0&0&0\\ 0&1&0&0&0&0\\ 0&0&1&0&0&0\\ 0&0&0&1&0&0\\ 0&0&0&0&1&0\\ 0&0&0&0&0&1},\quad
\widehat{N}_2=\smat{0&1&0&0&0&0\\ 1&0&1&0&0&0\\ 0&1&0&1&0&1\\ 0&0&1&0&1&0\\ 0&0&0&1&0&0\\ 0&0&1&0&0&0},\quad
\widehat{N}_3=\smat{0&0&1&0&0&0\\ 0&1&0&1&0&1\\ 1&0&2&0&1&0\\ 0&1&0&1&0&1\\ 0&0&1&0&0&0\\ 0&1&0&1&0&0}\nonumber\\
\widehat{N}_4&=\smat{0&0&0&1&0&0\\ 0&0&1&0&1&0\\ 0&1&0&1&0&1\\ 1&0&1&0&0&0\\ 0&1&0&0&0&0\\ 0&0&1&0&0&0},\quad
\widehat{N}_5=\smat{0&0&0&0&1&0\\ 0&0&0&1&0&0\\ 0&0&1&0&0&0\\ 0&1&0&0&0&0\\ 1&0&0&0&0&0\\ 0&0&0&0&0&1},\quad
\widehat{N}_6=\smat{0&0&0&0&0&1\\ 0&0&1&0&0&0\\ 0&1&0&1&0&0\\ 0&0&1&0&0&0\\ 0&0&0&0&0&1\\ 1&0&0&0&1&0}\qquad
\end{align}
where $N_4$ and $\widehat{N}_5$ encode the $\mathbb{Z}_2$ graph automorphism of $A_4$ and $E_6$. 
For type I theories, there exists an extended conformal chiral algebra with extended Virasoro characters encoded by the fundamental intertwiner $C=(n_{i 1}{}^{a})$. Accordingly, for the type I graphs, we have
\be
n_i=\sum_{a\in G} n_{i 1}{}^{a}\widehat N_{a},\qquad \widehat{\chi}_{r,a}(q)=\sum_{s=1}^{g-1} n_{s 1}{}^{a}{\chi}_{r,s}(q)\label{decomposition}
\ee
The first of these decompositions follows by specializing the compatibility condition in (\ref{FusionAlgebras}). For $E_6$:
 \be
 \arraycolsep=3.2pt
 C=(n_{i 1}{}^{a})=\smat{1&0&0&0&0&0\\ 0&1&0&0&0&0\\ 0&0&1&0&0&0\\ 0&0&0&1&0&1\\ 0&0&1&0&1&0\\ 
 0&1&0&1&0&0\\ 1&0&1&0&0&0\\ 0&1&0&0&0&1\\ 0&0&1&0&0&0\\ 0&0&0&1&0&0\\ 0&0&0&0&1&0};\qquad\quad
 \begin{array}{rl}
 \widehat{\chi}_{r,1}(q)&=\chi_{r,1}(q)+\chi_{r,7}(q)\\[4pt]
 \widehat{\chi}_{r,2}(q)&=\chi_{r,2}(q)+\chi_{r,6}(q)+\chi_{r,8}(q)\\[4pt]
 \widehat{\chi}_{r,3}(q)&=\chi_{r,3}(q)+\chi_{r,5}(q)+\chi_{r,7}(q)+\chi_{r,9}(q)\\[4pt]
 \widehat{\chi}_{r,4}(q)&=\chi_{r,4}(q)+\chi_{r,6}(q)+\chi_{r,10}(q)\\[4pt]
 \widehat{\chi}_{r,5}(q)&=\chi_{r,5}(q)+\chi_{r,11}(q)\\[4pt]
 \widehat{\chi}_{r,6}(q)&=\chi_{r,4}(q)+\chi_{r,8}(q)
 \end{array}\label{E6inter}
\ee

\subsection{Twisted conformal partition functions}
The twisted partition functions of the \ade theories are given by Petkova and Zuber~\cite{PetkovaZuber2001}. 
For the $A$ series, with $i,j,k\in A_{g-1}$, the Ocneanu fusion matrices are simply the Verlinde fusion matrices $N_i$ with conformal partition functions
\be
Z_i(q)=\sum_{j,k\in A_{g-1}} N_{ij}{}^k \chi_j(q)\chi_k(\bar q)
\ee 
More generally, for $(A_{g-2},G)$ cosets with $G$ of type I, the twisted partition functions are given by the sesquilinear forms
\be
Z^{(r,a,b,\kappa)}(q)=\!\!\!\!\sum_{(r',s'),(r'\!',s'\!')}\!\!\!
N_{r,r'}{}^{r'\!'}[P_{a,b}^{(\kappa)}]_{s's'\!'}
\chi_{r',s'}(q)\,\chi_{r'\!',s'\!'}(\overline{q})
=\!\!\!\sum_{(r',a'),(r'\!',a'\!')}\!\!\!\!
N_{r,r'}{}^{r'\!'}[\widehat{P}_{a,b}^{(\kappa)}]_{a'a'\!'}\,
\widehat{\chi}_{r',a'}(q)\,\widehat{\chi}_{r'\!',a'\!'}(\overline{q})
\ee
where the second form is new. Here $r,r',r''\in A_{g-2}$; $s,s',s''\in A_{g-1}$; $a,a',a'',b\in G$ and the {\em toric} and {\em extended toric} matrices are
\be
[P_{a,b}^{(\kappa)}]_{s's'\!'}
=\sum_{c\in T_\kappa} n_{s'a}{}^c\,n_{s'\!' b}{}^c,\qquad\ \ 
[\widehat{P}_{a,b}^{(\kappa)}]_{a'a'\!'}
=\sum_{c\in T_\kappa} \widehat{N}_{ac}{}^{a'}\,\widehat{N}_{bc}{}^{a'\!'},
\qquad \kappa=0,1\\[-4pt]
\ee
The label $\kappa=0,1$ is only needed for the $D_L$ ($L$ even) series and indicates the absence or presence of the $\mathbb{Z}_2$ automorphism. 
The {\em ambichiral vertices} (common to both left and right chiral subalgebras) are
\be
T_0=\begin{cases}
\{1,2,\ldots,L\},&G=A_L\\[-2pt]
\{1,3,5,\ldots,2\ell-1,2\ell\},&G=D_{2\ell}\\[-2pt]
\{1,5,6\},&G=E_6\\[-2pt]
\{1,7\},&G=E_8
\end{cases}\qquad\ \ 
T_1=\begin{cases}
\{2,4,\ldots,2\ell-2\},&G=D_{2\ell}\\
\;T_0,&\mbox{otherwise}\end{cases}
\ee

\subsection{Extended toric matrices and quantum symmetry}

\begin{figure}[tb]
\begin{center}
\ssmat{1&0&0&0&0&0\\ 0&0&0&0&0&0\\ 0&0&0&0&0&0\\ 0&0&0&0&0&0\\ 0&0&0&0&1&0\\ 0&0&0&0&0&1}\ \ 
\ssmat{0&1&0&0&0&0\\ 0&0&0&0&0&0\\ 0&0&0&0&0&0\\ 0&0&0&0&0&0\\ 0&0&0&1&0&0\\ 0&0&1&0&0&0}\ \ 
\ssmat{0&0&1&0&0&0\\ 0&0&0&0&0&0\\ 0&0&0&0&0&0\\ 0&0&0&0&0&0\\ 0&0&1&0&0&0\\ 0&1&0&1&0&0}\ \ 
\ssmat{0&0&0&1&0&0\\ 0&0&0&0&0&0\\ 0&0&0&0&0&0\\ 0&0&0&0&0&0\\ 0&1&0&0&0&0\\ 0&0&1&0&0&0}\ \ 
\ssmat{0&0&0&0&1&0\\ 0&0&0&0&0&0\\ 0&0&0&0&0&0\\ 0&0&0&0&0&0\\ 1&0&0&0&0&0\\ 0&0&0&0&0&1}\ \ 
\ssmat{0&0&0&0&0&1\\ 0&0&0&0&0&0\\ 0&0&0&0&0&0\\ 0&0&0&0&0&0\\ 0&0&0&0&0&1\\ 1&0&0&0&1&0}\nonumber\\[6pt]
 %%%%%%%%%%%%%%%%%%%%%%%%
\ssmat{0&0&0&0&0&0\\ 1&0&0&0&0&0\\ 0&0&0&0&0&1\\ 0&0&0&0&1&0\\ 0&0&0&0&0&0\\ 0&0&0&0&0&0}\ \ 
 %%%%%%%%%%%%%%%%%%%%%%%%%
\ssmat{0&0&0&0&0&0\\ 0&1&0&0&0&0\\ 0&0&1&0&0&0\\ 0&0&0&1&0&0\\ 0&0&0&0&0&0\\ 0&0&0&0&0&0}\ \ 
\ssmat{0&0&0&0&0&0\\ 0&0&1&0&0&0\\ 0&1&0&1&0&0\\ 0&0&1&0&0&0\\ 0&0&0&0&0&0\\ 0&0&0&0&0&0}\ \ 
\ssmat{0&0&0&0&0&0\\ 0&0&0&1&0&0\\ 0&0&1&0&0&0\\ 0&1&0&0&0&0\\ 0&0&0&0&0&0\\ 0&0&0&0&0&0}\ \ 
\ssmat{0&0&0&0&0&0\\ 0&0&0&0&1&0\\ 0&0&0&0&0&1\\ 1&0&0&0&0&0\\ 0&0&0&0&0&0\\ 0&0&0&0&0&0}\ \ 
 \ssmat{0&0&0&0&0&0\\ 0&0&0&0&0&1\\ 1&0&0&0&1&0\\ 0&0&0&0&0&1\\ 0&0&0&0&0&0\\ 0&0&0&0&0&0}\nonumber\\[6pt]
 %%%%%%%%%%%%%%%%%%%%%
 \ssmat{0&0&0&0&0&0\\ 0&0&0&0&0&1\\ 1&0&0&0&1&0\\ 0&0&0&0&0&1\\ 0&0&0&0&0&0\\ 0&0&0&0&0&0}\ \
\ssmat{0&0&0&0&0&0\\ 0&0&1&0&0&0\\ 0&1&0&1&0&0\\ 0&0&1&0&0&0\\ 0&0&0&0&0&0\\ 0&0&0&0&0&0}\ \ 
\ssmat{0&0&0&0&0&0\\ 0&1&0&1&0&0\\ 0&0&2&0&0&0\\ 0&1&0&1&0&0\\ 0&0&0&0&0&0\\ 0&0&0&0&0&0}\ \ 
\ssmat{0&0&0&0&0&0\\ 0&0&1&0&0&0\\ 0&1&0&1&0&0\\ 0&0&1&0&0&0\\ 0&0&0&0&0&0\\ 0&0&0&0&0&0}\ \ 
\ssmat{0&0&0&0&0&0\\ 0&0&0&0&0&1\\ 1&0&0&0&1&0\\ 0&0&0&0&0&1\\ 0&0&0&0&0&0\\ 0&0&0&0&0&0}\ \ 
 \ssmat{0&0&0&0&0&0\\ 1&0&0&0&1&0\\ 0&0&0&0&0&2\\ 1&0&0&0&1&0\\ 0&0&0&0&0&0\\ 0&0&0&0&0&0}\nonumber\\[6pt]
 %%%%%%%%%%%%%%%%%%
 \ssmat{0&0&0&0&0&0\\ 0&0&0&0&1&0\\ 0&0&0&0&0&1\\ 1&0&0&0&0&0\\ 0&0&0&0&0&0\\ 0&0&0&0&0&0}\ \ 
\ssmat{0&0&0&0&0&0\\ 0&0&0&1&0&0\\ 0&0&1&0&0&0\\ 0&1&0&0&0&0\\ 0&0&0&0&0&0\\ 0&0&0&0&0&0}\ \ 
 \ssmat{0&0&0&0&0&0\\ 0&0&1&0&0&0\\ 0&1&0&1&0&0\\ 0&0&1&0&0&0\\ 0&0&0&0&0&0\\ 0&0&0&0&0&0}\ \ 
\ssmat{0&0&0&0&0&0\\ 0&1&0&0&0&0\\ 0&0&1&0&0&0\\ 0&0&0&1&0&0\\ 0&0&0&0&0&0\\ 0&0&0&0&0&0}\ \ 
\ssmat{0&0&0&0&0&0\\ 1&0&0&0&0&0\\ 0&0&0&0&0&1\\ 0&0&0&0&1&0\\ 0&0&0&0&0&0\\ 0&0&0&0&0&0}\ \ 
\ssmat{0&0&0&0&0&0\\ 0&0&0&0&0&1\\ 1&0&0&0&1&0\\ 0&0&0&0&0&1\\ 0&0&0&0&0&0\\ 0&0&0&0&0&0}\\[6pt]
 %%%%%%%%%%%%%%%%%%%%%%%%%%%
\ssmat{0&0&0&0&1&0\\ 0&0&0&0&0&0\\ 0&0&0&0&0&0\\ 0&0&0&0&0&0\\ 1&0&0&0&0&0\\ 0&0&0&0&0&1}\ \ 
 \ssmat{0&0&0&1&0&0\\ 0&0&0&0&0&0\\ 0&0&0&0&0&0\\ 0&0&0&0&0&0\\ 0&1&0&0&0&0\\ 0&0&1&0&0&0}\ \ 
\ssmat{0&0&1&0&0&0\\ 0&0&0&0&0&0\\ 0&0&0&0&0&0\\ 0&0&0&0&0&0\\ 0&0&1&0&0&0\\ 0&1&0&1&0&0}\ \ 
\ssmat{0&1&0&0&0&0\\ 0&0&0&0&0&0\\ 0&0&0&0&0&0\\ 0&0&0&0&0&0\\ 0&0&0&1&0&0\\ 0&0&1&0&0&0}\ \ 
\ssmat{1&0&0&0&0&0\\ 0&0&0&0&0&0\\ 0&0&0&0&0&0\\ 0&0&0&0&0&0\\ 0&0&0&0&1&0\\ 0&0&0&0&0&1}\ \ 
\ssmat{0&0&0&0&0&1\\ 0&0&0&0&0&0\\ 0&0&0&0&0&0\\ 0&0&0&0&0&0\\ 0&0&0&0&0&1\\ 1&0&0&0&1&0}\nonumber\\[6pt]
%%%%%%%%%%%%%%%%%%%%%%%%%%%%
\ssmat{0&0&0&0&0&1\\ 0&0&0&0&0&0\\ 0&0&0&0&0&0\\ 0&0&0&0&0&0\\ 0&0&0&0&0&1\\ 1&0&0&0&1&0}\ \ 
\ssmat{0&0&1&0&0&0\\ 0&0&0&0&0&0\\ 0&0&0&0&0&0\\ 0&0&0&0&0&0\\ 0&0&1&0&0&0\\ 0&1&0&1&0&0}\ \ 
\ssmat{0&1&0&1&0&0\\ 0&0&0&0&0&0\\ 0&0&0&0&0&0\\ 0&0&0&0&0&0\\ 0&1&0&1&0&0\\ 0&0&2&0&0&0}\ \ 
\ssmat{0&0&1&0&0&0\\ 0&0&0&0&0&0\\ 0&0&0&0&0&0\\ 0&0&0&0&0&0\\ 0&0&1&0&0&0\\ 0&1&0&1&0&0}\ \ 
\ssmat{0&0&0&0&0&1\\ 0&0&0&0&0&0\\ 0&0&0&0&0&0\\ 0&0&0&0&0&0\\ 0&0&0&0&0&1\\ 1&0&0&0&1&0}\ \ 
\ssmat{1&0&0&0&1&0\\ 0&0&0&0&0&0\\ 0&0&0&0&0&0\\ 0&0&0&0&0&0\\ 1&0&0&0&1&0\\ 0&0&0&0&0&2}\nonumber
\end{center}
 %%%%%%%%%%%%%%%%%%%%%%%
 \vspace{-15pt}
\caption{The $6\times 6$ array of the extended toric matrices $\widehat{P}_{a,b}$ of $E_6$. The 12 basis matrices (\ref{basis}) are linearly independent. 
The other 24 matrices are given by the quantum symmetry (\ref{quant1}).
\label{Toric}}
\end{figure}

In the following, we fix $r=1$ and consider type 1 theories with $\kappa=0$. Taking $x=(a,b)$, the extended toric matrices can be written as
\be
\widehat{P}_{a,b}=(\widehat{N}_a)^T \widehat{N}_b,\qquad x=(a,b) 
\ee
where the superscript $T$ denotes transpose and, in accord with the tensor structure $G\otimes_{T_0}\!G$, the rows of the rectangular matrices $\widehat{N}_a$ are restricted to those labelled by the vertices of $T_0$.
Alternatively, working with the full $|G|\times |G|$ matrices, the extended toric matrices are given by the bilinear map
\be
\widehat{P}_{a,b}=(\widehat{N}_a,\widehat{N}_b)=\widehat{N}_a \widehat{P}\widehat{N}_b,\qquad 
(L,R)=L\widehat{P}R=L\otimes_{T_0}\!R,\qquad 
\widehat{P}=\widehat{P}_{1,1}=(I,I)
\ee
The projection $\widehat{P}=\widehat{P}_{1,1}$ joins the left and right chiral components through the ambichiral nodes $T_0$.

For $E_6$, the $6\times 6$ array of extended toric matrices $\widehat{P}_{a,b}$ is shown in Figure~\ref{Toric}. 
In accord with the quantum symmetry exhibited in Figure~\ref{Toric} and (\ref{quant1}), the 36 extended toric matrices lie in a convex cone $\mathbb K$ consisting of nonnegative-integer linear (conical) combinations of the (ordered) Hilbert basis~\cite{HilbertBasis}
\bea
\{\widehat{P}_\eta\}_{\eta=1}^{12}=\{\widehat{P}_{1},\widehat{P}_{2},\ldots,\widehat{P}_{12}\}=
\{\widehat{P}_{1,1},\widehat{P}_{2,1},\widehat{P}_{3,1},\widehat{P}_{4,1},\widehat{P}_{5,1},\widehat{P}_{6,1},\widehat{P}_{1,2},\widehat{P}_{2,2},\widehat{P}_{3,2},\widehat{P}_{4,2},\widehat{P}_{5,2}, \widehat{P}_{6,2}\}
\label{basis}
\eea
The 12 basis elements are in bijection with the vertices of the Ocneanu graph in Figure~\ref{E6OcneanuGraph}. 
They are conjugate to the 12 $E_6$ irreducible representations of $\text{Vir}\otimes \overline{\text{Vir}}$. 
The remaining 24 extended toric matrices are expressed as the conical linear combinations
\begin{subequations}
\label{quant1}
\begin{align}
&\hspace{2.5cm}\widehat{{P}}_{5,5}=\widehat{{P}}_{1,1},\qquad
\widehat{{P}}_{4,5}=\widehat{{P}}_{2,1}\label{quant1a},\qquad
\widehat{{P}}_{2,6}=\widehat{{P}}_{3,5}=\widehat{{P}}_{4,6}=\widehat{{P}}_{3,1}\\
&\hspace{2.5cm}\widehat{{P}}_{2,5}=\widehat{{P}}_{4,1},\qquad
\widehat{{P}}_{1,5}=\widehat{{P}}_{5,1},\qquad
\widehat{{P}}_{1,6}=\widehat{{P}}_{5,6}=\widehat{{P}}_{6,5}=\widehat{{P}}_{6,1}\\
&\hspace{2.5cm}\widehat{{P}}_{5,4}=\widehat{{P}}_{1,2},\qquad
\widehat{{P}}_{4,4}=\widehat{{P}}_{2,2},\qquad
\widehat{{P}}_{2,3}=\widehat{{P}}_{3,4}=\widehat{{P}}_{4,3}=\widehat{{P}}_{3,2}\\
&\hspace{2.5cm}\widehat{{P}}_{2,4}=\widehat{{P}}_{4,2},\qquad
\widehat{{P}}_{1,4}=\widehat{{P}}_{5,2},\qquad
\widehat{{P}}_{1,3}=\widehat{{P}}_{5,3}=\widehat{{P}}_{6,4}=\widehat{{P}}_{6,2}\label{quant1d}\\[1pt]
%\label{quantum2}
&\widehat{{P}}_{3,3}=\widehat{{P}}_{2,2}+\widehat{{P}}_{4,2},\qquad 
\widehat{{P}}_{3,6}=\widehat{{P}}_{2,1}+\widehat{{P}}_{4,1},\qquad
\widehat{{P}}_{6,3}=\widehat{{P}}_{1,2}+\widehat{{P}}_{5,2},\qquad
\widehat{{P}}_{6,6}=\widehat{{P}}_{1,1}+\widehat{{P}}_{5,1}
\end{align}
\end{subequations}
The Hilbert basis of extended toric matrices is unique up to these quantum symmetry equivalences.

It is useful to define the fusion product $\star$ of two extended toric matrices for type I theories by
\be
\widehat{P}_{a,a'}\star \widehat{P}_{b,b'}=(\widehat{N}_a,\widehat{N}_{a'})\star(\widehat{N}_b,\widehat{N}_{b'})
:=(\widehat{N}_a \widehat{N}_b,\widehat{N}_{a'} \widehat{N}_{b'})
=\sum_{c,c'\in G} \widehat{N}_{ab}{}^{c} \widehat{N}_{a'b'}{}^{c'} \widehat{P}_{c,c'},\quad a,a',b,b'\in G
\label{star}
\ee
This defines the double graph fusion algebra. As we discuss in the next subsection, it reduces to the Ocneanu algebra when the right side is expressed in terms of the Hilbert basis using the quantum symmetry (\ref{quant1}). 
In our notation, the left and right fundamentals are $\widehat{P}_{2,1}=(\widehat{N}_2,\widehat{N}_1)=(\widehat{N}_2,I)$ and 
$\widehat{P}_{1,2}=(\widehat{N}_1,\widehat{N}_2)=(I,\widehat{N}_2)$. 
For example, for $E_6$, straightforward calculation yields
\begin{align}
\widehat{P}_{3,2}\star \widehat{P}_{2,1}&=(\widehat{N}_3,\widehat{N}_2)\star(\widehat{N}_2,\widehat{N}_1)
=(\widehat{N}_3 \widehat{N}_2,\widehat{N}_2 \widehat{N}_1)=(\widehat{N}_2+\widehat{N}_4+\widehat{N}_6,\widehat{N}_2)
=\widehat{P}_{2,2}+\widehat{P}_{4,2}+\widehat{P}_{6,2}\\[6pt]
\widehat{P}_{1,2}\star \widehat{P}_{3,3}&=(\widehat{N}_1,\widehat{N}_2)\star(\widehat{N}_3,\widehat{N}_3)
=(\widehat{N}_1\widehat{N}_3,\widehat{N}_2\widehat{N}_3)=(\widehat{N}_3,\widehat{N}_2+\widehat{N}_4+\widehat{N}_6)\nonumber\\
&=(\widehat{N}_3,\widehat{N}_2)+(\widehat{N}_3,\widehat{N}_4)+(\widehat{N}_3,\widehat{N}_6)
=\widehat{P}_{3,2}+\widehat{P}_{3,4}+\widehat{P}_{3,6}
=\widehat{P}_{2,1}+2\widehat{P}_{3,2}+\widehat{P}_{4,1}
\end{align}

\subsection{$E_6$ Ocneanu algebra and quantum symmetry}

The Ocneanu graph $\widetilde{E}_6$ of $E_6$ is shown in Figure~\ref{E6OcneanuGraph}.
The twelve $12\times 12$ $E_6$ Ocneanu graph fusion matrices (nimreps) $\widetilde{N}_\eta$ with $\eta=1,2,\ldots,12$
are
\begin{align*}
&
\sssmat{
 1 & 0 & 0 & 0 & 0 & 0 & 0 & 0 & 0 & 0 & 0 & 0 \\
 0 & 1 & 0 & 0 & 0 & 0 & 0 & 0 & 0 & 0 & 0 & 0 \\
 0 & 0 & 1 & 0 & 0 & 0 & 0 & 0 & 0 & 0 & 0 & 0 \\
 0 & 0 & 0 & 1 & 0 & 0 & 0 & 0 & 0 & 0 & 0 & 0 \\
 0 & 0 & 0 & 0 & 1 & 0 & 0 & 0 & 0 & 0 & 0 & 0 \\
 0 & 0 & 0 & 0 & 0 & 1 & 0 & 0 & 0 & 0 & 0 & 0 \\
 0 & 0 & 0 & 0 & 0 & 0 & 1 & 0 & 0 & 0 & 0 & 0 \\
 0 & 0 & 0 & 0 & 0 & 0 & 0 & 1 & 0 & 0 & 0 & 0 \\
 0 & 0 & 0 & 0 & 0 & 0 & 0 & 0 & 1 & 0 & 0 & 0 \\
 0 & 0 & 0 & 0 & 0 & 0 & 0 & 0 & 0 & 1 & 0 & 0 \\
 0 & 0 & 0 & 0 & 0 & 0 & 0 & 0 & 0 & 0 & 1 & 0 \\
 0 & 0 & 0 & 0 & 0 & 0 & 0 & 0 & 0 & 0 & 0 & 1}\!
\sssmat{
 0 & 1 & 0 & 0 & 0 & 0 & 0 & 0 & 0 & 0 & 0 & 0 \\
 1 & 0 & 1 & 0 & 0 & 0 & 0 & 0 & 0 & 0 & 0 & 0 \\
 0 & 1 & 0 & 1 & 0 & 1 & 0 & 0 & 0 & 0 & 0 & 0 \\
 0 & 0 & 1 & 0 & 1 & 0 & 0 & 0 & 0 & 0 & 0 & 0 \\
 0 & 0 & 0 & 1 & 0 & 0 & 0 & 0 & 0 & 0 & 0 & 0 \\
 0 & 0 & 1 & 0 & 0 & 0 & 0 & 0 & 0 & 0 & 0 & 0 \\
 0 & 0 & 0 & 0 & 0 & 0 & 0 & 1 & 0 & 0 & 0 & 0 \\
 0 & 0 & 0 & 0 & 0 & 0 & 1 & 0 & 1 & 0 & 0 & 0 \\
 0 & 0 & 0 & 0 & 0 & 0 & 0 & 1 & 0 & 1 & 0 & 1 \\
 0 & 0 & 0 & 0 & 0 & 0 & 0 & 0 & 1 & 0 & 1 & 0 \\
 0 & 0 & 0 & 0 & 0 & 0 & 0 & 0 & 0 & 1 & 0 & 0 \\
 0 & 0 & 0 & 0 & 0 & 0 & 0 & 0 & 1 & 0 & 0 & 0}\!
\sssmat{
 0 & 0 & 1 & 0 & 0 & 0 & 0 & 0 & 0 & 0 & 0 & 0 \\
 0 & 1 & 0 & 1 & 0 & 1 & 0 & 0 & 0 & 0 & 0 & 0 \\
 1 & 0 & 2 & 0 & 1 & 0 & 0 & 0 & 0 & 0 & 0 & 0 \\
 0 & 1 & 0 & 1 & 0 & 1 & 0 & 0 & 0 & 0 & 0 & 0 \\
 0 & 0 & 1 & 0 & 0 & 0 & 0 & 0 & 0 & 0 & 0 & 0 \\
 0 & 1 & 0 & 1 & 0 & 0 & 0 & 0 & 0 & 0 & 0 & 0 \\
 0 & 0 & 0 & 0 & 0 & 0 & 0 & 0 & 1 & 0 & 0 & 0 \\
 0 & 0 & 0 & 0 & 0 & 0 & 0 & 1 & 0 & 1 & 0 & 1 \\
 0 & 0 & 0 & 0 & 0 & 0 & 1 & 0 & 2 & 0 & 1 & 0 \\
 0 & 0 & 0 & 0 & 0 & 0 & 0 & 1 & 0 & 1 & 0 & 1 \\
 0 & 0 & 0 & 0 & 0 & 0 & 0 & 0 & 1 & 0 & 0 & 0 \\
 0 & 0 & 0 & 0 & 0 & 0 & 0 & 1 & 0 & 1 & 0 & 0}\!
\sssmat{
 0 & 0 & 0 & 1 & 0 & 0 & 0 & 0 & 0 & 0 & 0 & 0 \\
 0 & 0 & 1 & 0 & 1 & 0 & 0 & 0 & 0 & 0 & 0 & 0 \\
 0 & 1 & 0 & 1 & 0 & 1 & 0 & 0 & 0 & 0 & 0 & 0 \\
 1 & 0 & 1 & 0 & 0 & 0 & 0 & 0 & 0 & 0 & 0 & 0 \\
 0 & 1 & 0 & 0 & 0 & 0 & 0 & 0 & 0 & 0 & 0 & 0 \\
 0 & 0 & 1 & 0 & 0 & 0 & 0 & 0 & 0 & 0 & 0 & 0 \\
 0 & 0 & 0 & 0 & 0 & 0 & 0 & 0 & 0 & 1 & 0 & 0 \\
 0 & 0 & 0 & 0 & 0 & 0 & 0 & 0 & 1 & 0 & 1 & 0 \\
 0 & 0 & 0 & 0 & 0 & 0 & 0 & 1 & 0 & 1 & 0 & 1 \\
 0 & 0 & 0 & 0 & 0 & 0 & 1 & 0 & 1 & 0 & 0 & 0 \\
 0 & 0 & 0 & 0 & 0 & 0 & 0 & 1 & 0 & 0 & 0 & 0 \\
 0 & 0 & 0 & 0 & 0 & 0 & 0 & 0 & 1 & 0 & 0 & 0}\!
\sssmat{
 0 & 0 & 0 & 0 & 1 & 0 & 0 & 0 & 0 & 0 & 0 & 0 \\
 0 & 0 & 0 & 1 & 0 & 0 & 0 & 0 & 0 & 0 & 0 & 0 \\
 0 & 0 & 1 & 0 & 0 & 0 & 0 & 0 & 0 & 0 & 0 & 0 \\
 0 & 1 & 0 & 0 & 0 & 0 & 0 & 0 & 0 & 0 & 0 & 0 \\
 1 & 0 & 0 & 0 & 0 & 0 & 0 & 0 & 0 & 0 & 0 & 0 \\
 0 & 0 & 0 & 0 & 0 & 1 & 0 & 0 & 0 & 0 & 0 & 0 \\
 0 & 0 & 0 & 0 & 0 & 0 & 0 & 0 & 0 & 0 & 1 & 0 \\
 0 & 0 & 0 & 0 & 0 & 0 & 0 & 0 & 0 & 1 & 0 & 0 \\
 0 & 0 & 0 & 0 & 0 & 0 & 0 & 0 & 1 & 0 & 0 & 0 \\
 0 & 0 & 0 & 0 & 0 & 0 & 0 & 1 & 0 & 0 & 0 & 0 \\
 0 & 0 & 0 & 0 & 0 & 0 & 1 & 0 & 0 & 0 & 0 & 0 \\
 0 & 0 & 0 & 0 & 0 & 0 & 0 & 0 & 0 & 0 & 0 & 1}\!
\sssmat{
 0 & 0 & 0 & 0 & 0 & 1 & 0 & 0 & 0 & 0 & 0 & 0 \\
 0 & 0 & 1 & 0 & 0 & 0 & 0 & 0 & 0 & 0 & 0 & 0 \\
 0 & 1 & 0 & 1 & 0 & 0 & 0 & 0 & 0 & 0 & 0 & 0 \\
 0 & 0 & 1 & 0 & 0 & 0 & 0 & 0 & 0 & 0 & 0 & 0 \\
 0 & 0 & 0 & 0 & 0 & 1 & 0 & 0 & 0 & 0 & 0 & 0 \\
 1 & 0 & 0 & 0 & 1 & 0 & 0 & 0 & 0 & 0 & 0 & 0 \\
 0 & 0 & 0 & 0 & 0 & 0 & 0 & 0 & 0 & 0 & 0 & 1 \\
 0 & 0 & 0 & 0 & 0 & 0 & 0 & 0 & 1 & 0 & 0 & 0 \\
 0 & 0 & 0 & 0 & 0 & 0 & 0 & 1 & 0 & 1 & 0 & 0 \\
 0 & 0 & 0 & 0 & 0 & 0 & 0 & 0 & 1 & 0 & 0 & 0 \\
 0 & 0 & 0 & 0 & 0 & 0 & 0 & 0 & 0 & 0 & 0 & 1 \\
 0 & 0 & 0 & 0 & 0 & 0 & 1 & 0 & 0 & 0 & 1 & 0}\qquad\qquad\\[4pt]
& \sssmat{
 0 & 0 & 0 & 0 & 0 & 0 & 1 & 0 & 0 & 0 & 0 & 0 \\
 0 & 0 & 0 & 0 & 0 & 0 & 0 & 1 & 0 & 0 & 0 & 0 \\
 0 & 0 & 0 & 0 & 0 & 0 & 0 & 0 & 1 & 0 & 0 & 0 \\
 0 & 0 & 0 & 0 & 0 & 0 & 0 & 0 & 0 & 1 & 0 & 0 \\
 0 & 0 & 0 & 0 & 0 & 0 & 0 & 0 & 0 & 0 & 1 & 0 \\
 0 & 0 & 0 & 0 & 0 & 0 & 0 & 0 & 0 & 0 & 0 & 1 \\
 1 & 0 & 0 & 0 & 0 & 0 & 0 & 0 & 0 & 0 & 0 & 1 \\
 0 & 1 & 0 & 0 & 0 & 0 & 0 & 0 & 1 & 0 & 0 & 0 \\
 0 & 0 & 1 & 0 & 0 & 0 & 0 & 1 & 0 & 1 & 0 & 0 \\
 0 & 0 & 0 & 1 & 0 & 0 & 0 & 0 & 1 & 0 & 0 & 0 \\
 0 & 0 & 0 & 0 & 1 & 0 & 0 & 0 & 0 & 0 & 0 & 1 \\
 0 & 0 & 0 & 0 & 0 & 1 & 1 & 0 & 0 & 0 & 1 & 0}\!
\sssmat{
 0 & 0 & 0 & 0 & 0 & 0 & 0 & 1 & 0 & 0 & 0 & 0 \\
 0 & 0 & 0 & 0 & 0 & 0 & 1 & 0 & 1 & 0 & 0 & 0 \\
 0 & 0 & 0 & 0 & 0 & 0 & 0 & 1 & 0 & 1 & 0 & 1 \\
 0 & 0 & 0 & 0 & 0 & 0 & 0 & 0 & 1 & 0 & 1 & 0 \\
 0 & 0 & 0 & 0 & 0 & 0 & 0 & 0 & 0 & 1 & 0 & 0 \\
 0 & 0 & 0 & 0 & 0 & 0 & 0 & 0 & 1 & 0 & 0 & 0 \\
 0 & 1 & 0 & 0 & 0 & 0 & 0 & 0 & 1 & 0 & 0 & 0 \\
 1 & 0 & 1 & 0 & 0 & 0 & 0 & 1 & 0 & 1 & 0 & 1 \\
 0 & 1 & 0 & 1 & 0 & 1 & 1 & 0 & 2 & 0 & 1 & 0 \\
 0 & 0 & 1 & 0 & 1 & 0 & 0 & 1 & 0 & 1 & 0 & 1 \\
 0 & 0 & 0 & 1 & 0 & 0 & 0 & 0 & 1 & 0 & 0 & 0 \\
 0 & 0 & 1 & 0 & 0 & 0 & 0 & 1 & 0 & 1 & 0 & 0}\!
\sssmat{
 0 & 0 & 0 & 0 & 0 & 0 & 0 & 0 & 1 & 0 & 0 & 0 \\
 0 & 0 & 0 & 0 & 0 & 0 & 0 & 1 & 0 & 1 & 0 & 1 \\
 0 & 0 & 0 & 0 & 0 & 0 & 1 & 0 & 2 & 0 & 1 & 0 \\
 0 & 0 & 0 & 0 & 0 & 0 & 0 & 1 & 0 & 1 & 0 & 1 \\
 0 & 0 & 0 & 0 & 0 & 0 & 0 & 0 & 1 & 0 & 0 & 0 \\
 0 & 0 & 0 & 0 & 0 & 0 & 0 & 1 & 0 & 1 & 0 & 0 \\
 0 & 0 & 1 & 0 & 0 & 0 & 0 & 1 & 0 & 1 & 0 & 0 \\
 0 & 1 & 0 & 1 & 0 & 1 & 1 & 0 & 2 & 0 & 1 & 0 \\
 1 & 0 & 2 & 0 & 1 & 0 & 0 & 2 & 0 & 2 & 0 & 2 \\
 0 & 1 & 0 & 1 & 0 & 1 & 1 & 0 & 2 & 0 & 1 & 0 \\
 0 & 0 & 1 & 0 & 0 & 0 & 0 & 1 & 0 & 1 & 0 & 0 \\
 0 & 1 & 0 & 1 & 0 & 0 & 0 & 0 & 2 & 0 & 0 & 0}\!
\sssmat{
 0 & 0 & 0 & 0 & 0 & 0 & 0 & 0 & 0 & 1 & 0 & 0 \\
 0 & 0 & 0 & 0 & 0 & 0 & 0 & 0 & 1 & 0 & 1 & 0 \\
 0 & 0 & 0 & 0 & 0 & 0 & 0 & 1 & 0 & 1 & 0 & 1 \\
 0 & 0 & 0 & 0 & 0 & 0 & 1 & 0 & 1 & 0 & 0 & 0 \\
 0 & 0 & 0 & 0 & 0 & 0 & 0 & 1 & 0 & 0 & 0 & 0 \\
 0 & 0 & 0 & 0 & 0 & 0 & 0 & 0 & 1 & 0 & 0 & 0 \\
 0 & 0 & 0 & 1 & 0 & 0 & 0 & 0 & 1 & 0 & 0 & 0 \\
 0 & 0 & 1 & 0 & 1 & 0 & 0 & 1 & 0 & 1 & 0 & 1 \\
 0 & 1 & 0 & 1 & 0 & 1 & 1 & 0 & 2 & 0 & 1 & 0 \\
 1 & 0 & 1 & 0 & 0 & 0 & 0 & 1 & 0 & 1 & 0 & 1 \\
 0 & 1 & 0 & 0 & 0 & 0 & 0 & 0 & 1 & 0 & 0 & 0 \\
 0 & 0 & 1 & 0 & 0 & 0 & 0 & 1 & 0 & 1 & 0 & 0}\!
\sssmat{
 0 & 0 & 0 & 0 & 0 & 0 & 0 & 0 & 0 & 0 & 1 & 0 \\
 0 & 0 & 0 & 0 & 0 & 0 & 0 & 0 & 0 & 1 & 0 & 0 \\
 0 & 0 & 0 & 0 & 0 & 0 & 0 & 0 & 1 & 0 & 0 & 0 \\
 0 & 0 & 0 & 0 & 0 & 0 & 0 & 1 & 0 & 0 & 0 & 0 \\
 0 & 0 & 0 & 0 & 0 & 0 & 1 & 0 & 0 & 0 & 0 & 0 \\
 0 & 0 & 0 & 0 & 0 & 0 & 0 & 0 & 0 & 0 & 0 & 1 \\
 0 & 0 & 0 & 0 & 1 & 0 & 0 & 0 & 0 & 0 & 0 & 1 \\
 0 & 0 & 0 & 1 & 0 & 0 & 0 & 0 & 1 & 0 & 0 & 0 \\
 0 & 0 & 1 & 0 & 0 & 0 & 0 & 1 & 0 & 1 & 0 & 0 \\
 0 & 1 & 0 & 0 & 0 & 0 & 0 & 0 & 1 & 0 & 0 & 0 \\
 1 & 0 & 0 & 0 & 0 & 0 & 0 & 0 & 0 & 0 & 0 & 1 \\
 0 & 0 & 0 & 0 & 0 & 1 & 1 & 0 & 0 & 0 & 1 & 0}\!
\sssmat{
 0 & 0 & 0 & 0 & 0 & 0 & 0 & 0 & 0 & 0 & 0 & 1 \\
 0 & 0 & 0 & 0 & 0 & 0 & 0 & 0 & 1 & 0 & 0 & 0 \\
 0 & 0 & 0 & 0 & 0 & 0 & 0 & 1 & 0 & 1 & 0 & 0 \\
 0 & 0 & 0 & 0 & 0 & 0 & 0 & 0 & 1 & 0 & 0 & 0 \\
 0 & 0 & 0 & 0 & 0 & 0 & 0 & 0 & 0 & 0 & 0 & 1 \\
 0 & 0 & 0 & 0 & 0 & 0 & 1 & 0 & 0 & 0 & 1 & 0 \\
 0 & 0 & 0 & 0 & 0 & 1 & 1 & 0 & 0 & 0 & 1 & 0 \\
 0 & 0 & 1 & 0 & 0 & 0 & 0 & 1 & 0 & 1 & 0 & 0 \\
 0 & 1 & 0 & 1 & 0 & 0 & 0 & 0 & 2 & 0 & 0 & 0 \\
 0 & 0 & 1 & 0 & 0 & 0 & 0 & 1 & 0 & 1 & 0 & 0 \\
 0 & 0 & 0 & 0 & 0 & 1 & 1 & 0 & 0 & 0 & 1 & 0 \\
 1 & 0 & 0 & 0 & 1 & 0 & 0 & 0 & 0 & 0 & 0 & 2}\qquad\qquad 
 \end{align*}
 \vspace{-20pt}
 \bea
 \label{OcneanuNimreps}
  \eea
 These fusion matrices satisfy the Ocneanu graph fusion algebra and  can be expressed compactly as  
 \bea
\widetilde{N}_\eta \widetilde{N}_\mu=\sum_{\nu=1}^{12} \widetilde{N}_{\eta\mu}{}^\nu\,\widetilde{N}_\nu,\qquad 
\widetilde{N}_\eta=\begin{cases}
\widehat{N}_\eta\oplus \widehat{N}_\eta,&\eta=1,2,\ldots,6\\
\Sigma\,\widetilde{N}_{\bar \eta}\,\Sigma,&\eta=7\\
\widetilde{N}_7 \widetilde{N}_{\eta-6},&\eta=8,9,\ldots,12
\end{cases}\qquad
\Sigma=\sssmat{
1 & 0 & 0 & 0 & 0 & 0 & 0 & 0 & 0 & 0 & 0 & 0 \\
 0 & 0 & 0 & 0 & 0 & 0 & 1 & 0 & 0 & 0 & 0 & 0 \\
 0 & 0 & 0 & 0 & 0 & 0 & 0 & 0 & 0 & 0 & 0 & 1 \\
 0 & 0 & 0 & 0 & 0 & 0 & 0 & 0 & 0 & 0 & 1 & 0 \\
 0 & 0 & 0 & 0 & 1 & 0 & 0 & 0 & 0 & 0 & 0 & 0 \\
 0 & 0 & 0 & 0 & 0 & 1 & 0 & 0 & 0 & 0 & 0 & 0 \\
 0 & 1 & 0 & 0 & 0 & 0 & 0 & 0 & 0 & 0 & 0 & 0 \\
 0 & 0 & 0 & 0 & 0 & 0 & 0 & 1 & 0 & 0 & 0 & 0 \\
 0 & 0 & 0 & 0 & 0 & 0 & 0 & 0 & 1 & 0 & 0 & 0 \\
 0 & 0 & 0 & 0 & 0 & 0 & 0 & 0 & 0 & 1 & 0 & 0 \\
 0 & 0 & 0 & 1 & 0 & 0 & 0 & 0 & 0 & 0 & 0 & 0 \\
 0 & 0 & 1 & 0 & 0 & 0 & 0 & 0 & 0 & 0 & 0 & 0}&
\eea
The $\mathbb{Z}_2$ chiral conjugation $\Sigma$ implements the interchanges $\eta\leftrightarrow \bar \eta$ with $2\leftrightarrow 7,3\leftrightarrow 12,4\leftrightarrow 11$ and $\bar\eta=\eta$ for $\eta=1,5,6,8,9,10$.

It is readily verified that these basis toric matrices (\ref{basis}) satisfy the Ocneanu algebra
\bea
\widehat{P}_\eta\star\widehat{P}_\mu=\sum_{\nu=1}^{12}\widetilde{N}_{\eta\mu}{}^\nu \widehat{P}_\nu,\qquad \eta,\mu=1,2,\ldots,12
\eea
Due to the chiral action of the $\star$ fusion product, which multiplies left with left chiral components and right with right chiral components, this realization of the Ocneanu algebra does not yield simple matrix product representations of the algebra. 
If we now combine left and right chiral components we obtain
\bea
\widehat{P}_\eta\star\widehat{P}_{\,\bar \mu}=\sum_{\nu=1}^{12}\widetilde{N}_{\eta\bar \mu}{}^\nu \widehat{P}_\nu,\qquad \eta,\mu=1,2,\ldots,12
\eea
For $1\le \eta,\mu\le 6$, $\widehat{P}_\eta\star\widehat{P}_{\,\bar \mu}=\widehat{P}_{\eta,\bar \mu}$ and restricting to $3\le \mu\le 6$ in the right side above reproduces the quantum symmetry relations (\ref{quant1}).

%%%%%%%%%%%%%%%%%%%%%%%%%%%%
%
\section{\ade integrable seams}\label{sec:Ocneanu}
%
%%%%%%%%%%%%%%%%%%%%%%%%%%%%

In this section we use bold face symbols ${\nn s}, \widehat{\mathbf N}_a, \overline{\widehat{\mathbf N}}_b$ to denote integrable seams with system size $M$ (see Figure~\ref{TMs}). This distinguishes them from their associated nimreps such as $n_s,\widehat{N}_a, \widehat{N}_b$. The nimreps give the adjacency conditions on the horizontal bonds of the vertical integrable seams $\nn s, \widehat{\mathbf{N}}_a,\overline{\widehat{\mathbf{N}}}_b$.

\subsection{Construction of $s$-type seams}\label{sec:Braid}

The fundamental braid seams ${\mathbf B},\overline{\mathbf B}$ are built analogously to the transfer matrix $\T(u)$ in (\ref{Tv}) but with the face weights replaced by the complex weights (\ref{braid}) and (\ref{invbraid}) respectively. These are just the braid limits of the commuting family $\T(u)$, suitably normalized, so ${\mathbf B}$ and $\overline{\mathbf B}$ commute with $\T(u)$ and with each other. Since $\overline{\mathbf B}={\mathbf B}^T$, ${\mathbf B}$ is Hermitian. It also has the same real eigenvalues as $G$. The fused $s$-type braid seams $\nn s, \nb s$ are defined recursively by the analog of (\ref{nRecursion})
\begin{subequations}
\begin{align}
&\nn 0=0,\qquad \nn 1=I,\qquad \nn 2={\mathbf B},\qquad \nn s\nn 2=\nn{s-1}+\nn{s+1}\\
&\nb 0=0,\qquad \nb 1=I,\qquad \nb 2=\overline{\mathbf B},\qquad \nb s\nb 2=\nb{s-1}+\nb{s+1}
\end{align}
\end{subequations}
Alternatively, these braid seams are constructed~\cite{PZ94} using Wenzl-Jones projectors to build the fused transfer matrices $\T^{1,s}(u)$ and then taking the braid limits. The fused braid matrices $\nn s, \nb s$ share the same real eigenvalues which are a subset of the eigenvalues of the Verlinde matrices $N_s$. The $\nn s, \nb s$ integrable seams are called $s$-type seams because they are associated with the $s$ label in the $(r,s)$ Kac labelling of the irreps of $\mbox{Vir}$.

If $G=A_{g-1}$ then, for arbitrary system size $M$, $\NN s=\Nb s=\nn s$ give matrix representations of the Verlinde algebra
\be 
\NN i\,\NN j=\sum_{k=1}^{g-1} N_{ij}{}^k\,\NN k
\ee
In these cases, the Ocneanu algebra reduces to the Verlinde algebra.

\subsection{Construction of $a$-type and $b$-type seams}
For $D$ and $E$ theories, we now extend the decomposition (\ref{decomposition}) to relate fused seams of types $s$, $a$ and $b$
\be
\nn s=\sum_{a\in G} n_{s 1}{}^{a}\,{\widehat{\mathbf N}_a},\qquad \nb s=\sum_{b\in G} n_{s 1}{}^{b}\,{\overline{\widehat{\mathbf N}}_b},\qquad s=1,2,\ldots,g-1\label{CNhat}
\ee
The face weights of the fused seams ${\widehat{\mathbf N}_a}, {\overline{\widehat{\mathbf N}}_b}$ can be calculated systematically as explained in \cite{ChuiEtAl2003}. However, the form of these face weights is not unique. Indeed, they depend on a choice of gauge. Here we will work in a different gauge to \cite{ChuiEtAl2003}. For arbitrary system size $M$, the fused seam matrices constructed here provide matrix representations of the same  fused adjacency and graph fusion algebras as in (\ref{FusionAlgebras}) with the same structure constants:
\begin{subequations}
\begin{align}
&\nn i\,\nn j=\sum_{k=1}^{g-1} N_{ij}{}^k\,\nn k,\qquad 
{\widehat{\mathbf N}_a}\,{\widehat{\mathbf N}_b}=\sum_{c\in G} \widehat{N}_{ab}{}^c\,{\widehat{\mathbf N}_c},\qquad 
\nn i\,{\widehat{\mathbf N}_a}=\sum_{b\in G} n_{ia}{}^b\,{\widehat{\mathbf N}_b}\\
&\nb i\,\nb j=\sum_{k=1}^{g-1} N_{ij}{}^k\,\nb k,\qquad 
{\overline{\widehat{\mathbf N}}_a}\,{\overline{\widehat{\mathbf N}}_b}=\sum_{c\in G} \widehat{N}_{ab}{}^c\,{\overline{\widehat{\mathbf N}}_c},\qquad 
\nb i\,{\overline{\widehat{\mathbf N}}_b}=\sum_{c\in G} n_{ib}{}^c\,{\overline{\widehat{\mathbf N}}_c}
\end{align}
\label{seamFusionAlgebras}
\end{subequations}

For $E_6$, the rectangular intertwiner $C=(n_{s1}{}^b)$ in (\ref{E6inter}) admits a generalized left inverse $C_\text{\scriptsize left}^{-1}$ such that $C_\text{\scriptsize left}^{-1}C=I$.
So (\ref{CNhat}) can be inverted to give the graph fusion seam matrices as
\begin{subequations}
\begin{align}
{\widehat{\mathbf N}_a}=\nn a,\ \ a=1,2,3;\quad\ \  {\widehat{\mathbf N}_4}=\nn 6-\nn 2,\quad
{\widehat{\mathbf N}_5}=\nn 5-\nn 3,\quad {\widehat{\mathbf N}_6}=\nn 2+\nn 4-\nn 6\\[4pt]
{\overline{\widehat{\mathbf N}}_b}=\nb b,\ \ b=1,2,3;\ \ \quad{\overline{\widehat{\mathbf N}}_4}=\nb 6-\nb 2,\quad
{\overline{\widehat{\mathbf N}}_5}=\nb 5-\nb 3,\quad {\overline{\widehat{\mathbf N}}_6}=\nb 2+\nb 4-\nb 6
\end{align}
\end{subequations}
With these, it is readily verified using Mathematica~\cite{Wolfram} that, for even system sizes up to $M=12$, all of the relations (\ref{seamFusionAlgebras}) are satisfied and that all of these matrices commute. In fact these relations hold for all $M$ by construction.

The local face weights of ${\widehat{\mathbf N}_a}=\nn a$ with $a=1,2,3$ follow from \cite{PZ94}. For $a=3$, these face weights involve bond variables. 
For ${\widehat{\mathbf N}_a}$ with $a=4,5,6$, we find that the allowed face weights are given by
\begin{align}
\hspace{-1.5cm}{\widehat{\mathbf N}_4}:\quad &\yface 1432=\yface 2341=\yface 4123=\yface 3214=(\sqrt{3}-1)^{\frac 14},\ \ \ \ \yface 2323=\yface 4343=2^{-\frac 14}e^{3\pi i/4}\nonumber\\[4pt]
&\yface 4125=\yface 5214=\yface 2541=\yface 1452=(2-\sqrt{3})^{\frac 14},\ \ \ \  \yface 3232=\yface 3434=2^{-\frac 14}e^{-3\pi i/4}\nonumber\\[4pt]
&\yface 2343=\yface 3432=\yface 3234=\yface 4323=\big(\tfrac 12(7-4\sqrt{3})\big)^{\frac 14},\qquad \yface 3636=e^{-5\pi i/12}(2-\sqrt{3})^{\frac 14}\\[4pt]
&\yface 2363=\yface 3632=\yface 3236=\yface 6323=
   \yface 3436=\yface 6343=\yface 4363=\yface 3634=\big(\tfrac 12(\sqrt{3}-1)\big)^{\frac 14}\nonumber\\[4pt]
&\yface 2543=\yface 3452=\yface 5234=\yface 4325=-(\sqrt{3}-1)^{\frac 14},\qquad\quad\   \yface 6363=e^{5\pi i/12}(2-\sqrt{3})^{\frac 14} \nonumber
\end{align}
\begin{align}
\hspace{-.5cm}{\widehat{\mathbf N}_5}:\quad 
&\yface 1542=\yface 2451=\yface 5124=\yface 4215=\yface 3366=\yface 6633=1\hspace{4.95cm}\\[4pt]
&\yface 3342=\yface 2433=\yface 4233=\yface 3324=1\nonumber
\end{align}
\begin{align}
\hspace{-1.65cm}\;{\widehat{\mathbf N}_6}:\quad 
&\yface 1632=\yface 2361=\yface 3216=\yface 6123=\yface 3456=\yface 6543=\yface 4365=\yface 5634=1\hspace{-.0cm}\\[4pt]
&\yface 3432=\yface 2343=\yface 3234=\yface 4323=\yface 2323=\yface 3232=2^{-\frac 14},\ \ \ \ \ \yface 3434=\yface 4343=-2^{-\frac 14} \nonumber
\end{align}
The face weights not explicitly listed vanish. Here $\widehat{\mathbf N}_5=\boldsymbol\sigma$ is the ${\mathbb Z}_2$ $E_6$ graph automorphism seam. For ${\overline{\widehat{\mathbf N}}_a}$ with $a=4,5,6$, the face weights are obtained by complex conjugation. Reflection in the horizontal implements complex conjugation. 
We note that $\widehat{\mathbf N}_a=\overline{\widehat{\mathbf N}}_a$ is real for the ambichiral vertices $a=1,5,6$. 

\subsection{Ocneanu algebra of seams and quantum symmetry}

For even system sizes up to $M=12$, it is now readily confirmed that the $E_6$ integrable seams satisfy the double graph fusion algebra
\bea
\widehat{{\mathbf P}}_{a,b}
=\widetilde{\mathbf N}_{a,b}={\widehat{\mathbf N}_a}\, {\overline{\widehat{\mathbf N}}_b},
\qquad\widetilde{\mathbf N}_{ab}\,\widetilde{\mathbf N}_{a'b'}=\sum_{a'\!',b'\!'=1}^6 \widehat{N}_{aa'}{}^{a'\!'} \widehat{N}_{bb'}{}^{b'\!'}\widetilde{\mathbf N}_{a'\!'b'\!'}
\label{PFactor}
\eea
%\end{subequations}
with $\overline{\widehat{{\mathbf P}}}_{a,b}=\widehat{{\mathbf P}}_{b,a}$. Alternatively, using ordered sets, we can define
\bea
\{\widehat{\mathbf P}_\eta\}_{\eta=1}^{12}=\{\widehat{\mathbf P}_1,\widehat{\mathbf P}_2,\ldots,\widehat{\mathbf P}_{12}\}
=\{\widehat{\mathbf P}_{1,1},\widehat{\mathbf P}_{2,1},\ldots,\widehat{\mathbf P}_{6,1},\widehat{\mathbf P}_{1,2},\widehat{\mathbf P}_{2,2},\ldots,\widehat{\mathbf P}_{6,2}\}
\label{hatPx}
\eea
with $\overline{\widehat{\mathbf P}}_\eta=\widehat{\mathbf P}_{\bar \eta}$. For $M\le 12$, it is verified that these integrable seams satisfy the Ocneanu algebra
\bea
\widehat{\mathbf P}_\eta\widehat{\mathbf P}_\mu=\sum_{\nu=1}^{12}\widetilde{N}_{\eta\mu}{}^\nu \widehat{\mathbf P}_\nu,\qquad \eta,\mu=1,2,\ldots,12
\label{PxOcneanu}
\eea
In this realization involving fusion of seams, the fusion product $\star$ is replaced with the usual matrix product. 
In a sense, the decomposition of fusion products of seams on the lattice is the analog of the local operator product expansion in CFT.

\begin{figure}[htb]
\begin{center}
\psset{unit=0.8cm}
\qquad\begin{pspicture}[shift=-.40](0,0)(4,7.25)
\psline[linewidth=1.pt,linecolor=red](-.5,1)(-.5,5)
\multirput(0,0)(0,-2){3}{\psline[linewidth=1.pt,linecolor=red](-.5,5)(1.2,6.4)}
\psline[linewidth=1.pt,linecolor=blue](3.5,1)(3.5,5)
\multirput(0,0)(0,-2){3}{\psline[linewidth=1.pt,linecolor=blue](3.5,5)(1.2,6.4)}
\psline[linewidth=1.pt,linecolor=blue,linestyle=dashed, dash=2pt 2pt](1.77,-.5)(1.77,3.5)
\psline[linewidth=1.pt,linecolor=red,linestyle=dashed, dash=2pt 2pt](1.83,-.5)(1.83,3.5)
\multirput(0,0)(0,-2){3}{\psline[linewidth=1.pt,linecolor=red,linestyle=dashed, dash=2pt 2pt](3.5,5)(1.8,3.6)}
\multirput(0,0)(0,-2){3}{\psline[linewidth=1.pt,linecolor=blue,linestyle=dashed, dash=2pt 2pt](-.5,5)(1.8,3.6)}
\multirput(-.5,1)(0,2){3}{\pscircle[linewidth=.5pt,fillstyle=solid,fillcolor=black](0,0){.13}}
\multirput(1.2,2.4)(0,2){3}{\pscircle[linewidth=.5pt,fillstyle=solid,fillcolor=black](0,0){.13}}
\multirput(3.5,1)(0,2){3}{\pscircle[linewidth=.5pt,fillstyle=solid,fillcolor=lightyellow](0,0){.13}}
\multirput(1.8,-.4)(0,2){3}{\pscircle[linewidth=.5pt,fillstyle=solid,fillcolor=lightyellow](0,0){.13}}
\rput(1.3,6.8){$1$}
\rput(-.9,5){$2$}
\rput(-.9,3){$3$}
\rput(-.9,1){$4$}
\rput(1.3,2.8){$5$}
\rput(1.3,4.8){$6$}
\rput(4.,5){$7$}
\rput(2.2,3.3){$8$}
\rput(2.2,1.3){$9$}
\rput(2.2,-.7){$10$}
\rput(4.,1){$11$}
\rput(4.,3){$12$}
\rput(-2.5,3.){\large $\widetilde{E}_6$:}
\end{pspicture}\qquad\qquad\qquad
%E_6
%\psset{unit=1.2cm}
\raisebox{0cm}{\begin{pspicture}[shift=-.40](0,0)(4,7.25)
\psline[linewidth=1.pt,linecolor=red](-.5,1)(-.5,5)
\multirput(0,0)(0,-2){3}{\psline[linewidth=1.pt,linecolor=red](-.5,5)(1.2,6.4)}
\psline[linewidth=1.pt,linecolor=blue](3.5,1)(3.5,5)
\multirput(0,0)(0,-2){3}{\psline[linewidth=1.pt,linecolor=blue](3.5,5)(1.2,6.4)}
\psline[linewidth=1.pt,linecolor=blue,linestyle=dashed, dash=2pt 2pt](1.77,-.5)(1.77,3.5)
\psline[linewidth=1.pt,linecolor=red,linestyle=dashed, dash=2pt 2pt](1.83,-.5)(1.83,3.5)
\multirput(0,0)(0,-2){3}{\psline[linewidth=1.pt,linecolor=red,linestyle=dashed, dash=2pt 2pt](3.5,5)(1.8,3.6)}
\multirput(0,0)(0,-2){3}{\psline[linewidth=1.pt,linecolor=blue,linestyle=dashed, dash=2pt 2pt](-.5,5)(1.8,3.6)}
\multirput(-.5,1)(0,2){3}{\pscircle[linewidth=.5pt,fillstyle=solid,fillcolor=black](0,0){.13}}
\multirput(1.2,2.4)(0,2){3}{\pscircle[linewidth=.5pt,fillstyle=solid,fillcolor=black](0,0){.13}}
\multirput(3.5,1)(0,2){3}{\pscircle[linewidth=.5pt,fillstyle=solid,fillcolor=lightyellow](0,0){.13}}
\multirput(1.8,-.4)(0,2){3}{\pscircle[linewidth=.5pt,fillstyle=solid,fillcolor=lightyellow](0,0){.13}}
\rput(1.3,6.9){\small $\widehat{{\mathbf P}}_{1,1}$}
\rput(1.3,4.9){\small $\widehat{{\mathbf P}}_{6,1}$}
\rput(1.3,2.9){\small $\widehat{{\mathbf P}}_{5,1}$}
\rput(-1.1,5){\small $\widehat{{\mathbf P}}_{2,1}$}
\rput(-1.1,3){\small $\widehat{{\mathbf P}}_{3,1}$}
\rput(-1.1,1){\small $\widehat{{\mathbf P}}_{4,1}$}
\rput(2.4,3.3){\small $\widehat{{\mathbf P}}_{2,2}$}
\rput(2.4,1.3){\small $\widehat{{\mathbf P}}_{3,2}$}
\rput(2.4,-.7){\small $\widehat{{\mathbf P}}_{4,2}$}
\rput(4.2,5){\small $\widehat{{\mathbf P}}_{1,2}$}
\rput(4.2,3){\small $\widehat{{\mathbf P}}_{6,2}$}
\rput(4.2,1){\small $\widehat{{\mathbf P}}_{5,2}$}
%\rput(.2,-.6){$\widetilde{E}_6$}
\end{pspicture}}
\vspace{.3cm}
\end{center}
\caption{Ocneanu fusion graph $\widetilde{E}_6$ of $E_6$ with 12 vertices and two alternative labellings of nodes.
The 12 vertices are in bijection with 
(i) the Hilbert basis $\widehat{P}_\eta$ (\ref{basis}) with $\eta=1,2,\ldots,12$ and
(ii) the linearly independent integrable seams $\widehat{{\mathbf P}}_{\eta}$ with $\eta=1,2,\ldots,12$ as in (\ref{hatPx}). The red and blue lines (solid or dashed) show the action of the fundamental seams $\widehat{{\mathbf P}}_{2}=\widehat{{\mathbf P}}_{2,1}$ with $\widehat{{\mathbf P}}_{7}=\widehat{{\mathbf P}}_{\bar 2}=\widehat{{\mathbf P}}_{1,2}$ respectively. This action interlocks four copies of the $E_6$ graph. The seams satisfy 
$\overline{\widehat{{\mathbf P}}}_{\eta}={\widehat{{\mathbf P}}}_{\bar \eta}$ where $\widehat{\mathbf P}_a=\widehat{\mathbf P}_{a,1}$ and 
$\widehat{\mathbf P}_{a+6}=\widehat{\mathbf P}_{a,2}$ with $a=1,2,\ldots,6$.
 \label{E6OcneanuGraph}}
\end{figure}

In accord with the quantum symmetry (\ref{quant1}) exhibited by the 36 ``extended toric seams'' (see Figure~\ref{Toric}), the Hilbert basis of composite seams (\ref{hatPx}) 
is in bijection with the 12 nodes of the Ocneanu graph in Figure~\ref{E6OcneanuGraph}. 
The remaining 24 ``extended toric seams'' are given by the conical linear combinations
\begin{subequations}
\label{quantum1}
\begin{align}
&\hspace{2.5cm}\widehat{{\mathbf P}}_{5,5}=\widehat{{\mathbf P}}_{1,1},\qquad
\widehat{{\mathbf P}}_{4,5}=\widehat{{\mathbf P}}_{2,1},\qquad
\widehat{{\mathbf P}}_{2,6}=\widehat{{\mathbf P}}_{3,5}=\widehat{{\mathbf P}}_{4,6}=\widehat{{\mathbf P}}_{3,1}\\
&\hspace{2.5cm}\widehat{{\mathbf P}}_{2,5}=\widehat{{\mathbf P}}_{4,1},\qquad
\widehat{{\mathbf P}}_{1,5}=\widehat{{\mathbf P}}_{5,1},\qquad
\widehat{{\mathbf P}}_{1,6}=\widehat{{\mathbf P}}_{5,6}=\widehat{{\mathbf P}}_{6,5}=\widehat{{\mathbf P}}_{6,1}\\
&\hspace{2.5cm}\widehat{{\mathbf P}}_{5,4}=\widehat{{\mathbf P}}_{1,2},\qquad
\widehat{{\mathbf P}}_{4,4}=\widehat{{\mathbf P}}_{2,2},\qquad
\widehat{{\mathbf P}}_{2,3}=\widehat{{\mathbf P}}_{3,4}=\widehat{{\mathbf P}}_{4,3}=\widehat{{\mathbf P}}_{3,2}\\
&\hspace{2.5cm}\widehat{{\mathbf P}}_{2,4}=\widehat{{\mathbf P}}_{4,2},\qquad
\widehat{{\mathbf P}}_{1,4}=\widehat{{\mathbf P}}_{5,2},\qquad
\widehat{{\mathbf P}}_{1,3}=\widehat{{\mathbf P}}_{5,3}=\widehat{{\mathbf P}}_{6,4}=\widehat{{\mathbf P}}_{6,2}\\[1pt]
%\label{quantum2}
&\widehat{{\mathbf P}}_{3,3}=\widehat{{\mathbf P}}_{2,2}+\widehat{{\mathbf P}}_{4,2},\qquad 
\widehat{{\mathbf P}}_{3,6}=\widehat{{\mathbf P}}_{2,1}+\widehat{{\mathbf P}}_{4,1},\qquad
\widehat{{\mathbf P}}_{6,3}=\widehat{{\mathbf P}}_{1,2}+\widehat{{\mathbf P}}_{5,2},\qquad
\widehat{{\mathbf P}}_{6,6}=\widehat{{\mathbf P}}_{1,1}+\widehat{{\mathbf P}}_{5,1}
\end{align}
\end{subequations}
where again this was checked in Mathematica out to even system sizes $M=12$. 
Replacing $b$ with $\bar b$ in (\ref{PxOcneanu}) gives
\bea
{\widehat{\mathbf P}_{\eta,\mu}=\widehat{{\mathbf N}}_{\eta}\,\overline{\widehat{{\mathbf N}}}_{\mu}=\widehat{{\mathbf N}}_{\eta}\,{\widehat{{\mathbf N}}_{\bar \mu}}
=\sum_{\nu=1}^{12} \widetilde{N}_{\eta\bar \mu}{}^{\,\nu}\, \widehat{{\mathbf P}}_{\nu},\qquad \eta,\mu=1,2,\ldots,12}
\eea
For $1\le \eta\le 6$ and $3\le \mu\le 6$, this reproduces the quantum symmetry (\ref{quantum1}). Using this to express the right side of the double graph fusion algebra (\ref{PFactor}) in basis seams (\ref{hatPx}) reproduces the Ocneanu algebra (\ref{PxOcneanu}).

In general, the representations may not be faithful for small values of $M$. For $E_6$, the representations are not faithful for $M<6$. 
The quantum symmetry encodes the rules for fusing left and right chiral seams. 
The $E_6$ Cayley table of the Ocneanu algebra of seams is shown in Figure~\ref{E6Cayley}.

\begin{figure}[htb]
\begin{center}
\small
\mbox{}\hspace{-0pt}\mbox{}
\arraycolsep=1.0pt
$\begin{array}{c||c|c|c|c|c|c||c|c|c|c|c|c|}
%\hline
&1&2&3&4&5&6&7&8&9&10&11&12 \rule{0pt}{6pt}\\
\hline\hline
1&1&2&3&4&5&6&7&8&9&10&11&12 \rule{0pt}{6pt}\\
\hline
2&2&1\!\!+\!\!3&2\!\!+\!\!4\!\!+\!\!6&3\!\!+\!\!5&4&3&8&7\!\!+\!\!9&8\!\!+\!\!10\!\!+\!\!12&9\!\!+\!\!11&10&9 \rule{0pt}{6pt}\\
\hline
3&3&2\!\!+\!\!4\!\!+\!\!6&1\!\!+\!\!2(3)\!\!+\!\!5&2\!\!+\!\!4\!\!+\!\!6&3&2\!\!+\!\!4&9&8\!\!+\!\!10\!\!+\!\!12&7\!\!+\!\!2(9)\!\!+\!\!11&8\!\!+\!\!10\!\!+\!\!12&9&8\!\!+\!\!10 \rule{0pt}{6pt}\\
\hline
4&4&3\!\!+\!\!5&2\!\!+\!\!4\!\!+\!\!6&1\!\!+\!\!3&2&3&10&9\!\!+\!\!11&8\!\!+\!\!10\!\!+\!\!12&7\!\!+\!\!9&8&9 \rule{0pt}{6pt}\\
\hline
5&5&4&3&2&1&6&11&10&9&8&7&12 \rule{0pt}{6pt}\\
\hline
6&6&3&2\!\!+\!\!4&3&6&1\!\!+\!\!5&12&9&8\!\!+\!\!10&9&12&7\!\!+\!\!11 \rule{0pt}{6pt}\\
\hline\hline
7&7&8&9&10&11&12&1\!\!+\!\!12&2\!\!+\!\!9&3\!\!+\!\!8\!\!+\!\!10&4\!\!+\!\!9&5\!\!+\!\!12&6\!\!+\!\!7\!\!+\!\!11 \rule{0pt}{6pt}\\
\hline
8&8&7\!\!+\!\!9&8\!\!+\!\!10\!\!+\!\!12&9\!\!+\!\!11&10&9&2\!\!+\!\!9&
%11\!\!+\!\!8\!\!+\!\!3\!\!+\!\!10\!\!+\!\!12&
\mbox{$1\!\!+\!\!3\!\!+\!\!8$\rule[-3pt]{0pt}{10pt}}\atop\mbox{$\!+10\!\!+\!\!12$\rule[-3pt]{0pt}{8pt}}&
%7\!\!+\!\!2\!\!+\!\!2(9)\!\!+\!\!4\!\!+\!\!11\!\!+\!\!6
\mbox{$2\!\!+\!\!4\!\!+\!\!6\!\!+\!\!7$\rule[-3pt]{0pt}{10pt}}\atop\mbox{$\!+2(9)\!\!+\!\!11$\rule[-3pt]{0pt}{8pt}}&
%8\!\!+\!\!3\!\!+\!\!10\!\!+\!\!5\!\!+\!\!12
\mbox{$3\!\!+\!\!5\!\!+\!\!8$\rule[-3pt]{0pt}{10pt}}\atop\mbox{$+10\!\!+\!\!12$\rule[-3pt]{0pt}{8pt}}
&
4\!\!+\!\!9&3\!\!+\!\!8\!\!+\!\!10 \rule{0pt}{6pt}\\
\hline
9&9&8\!\!+\!\!10\!\!+\!\!12&7\!\!+\!\!2(9)\!\!+\!\!11&8\!\!+\!\!10\!\!+\!\!12&9&8\!\!+\!\!10&3\!\!+\!\!8\!\!+\!\!10&
\mbox{$2\!\!+\!\!4\!\!+\!\!6\!\!+\!\!7$\rule[-3pt]{0pt}{10pt}}\atop\mbox{$\!+2(9)\!\!+\!\!11$\rule[-3pt]{0pt}{8pt}}&
\mbox{$1\!\!+\!\!2(3)\!\!+\!\!5\!\!+2(8)$\rule[-3pt]{0pt}{10pt}}\atop\mbox{$\!+2(10)\!\!+\!\!2(12)\!$\rule[-3pt]{0pt}{8pt}}&
%7\!\!+\!\!2\!\!+\!\!2(9)\!\!+\!\!4\!\!+\!\!11\!\!+\!\!6
%\mbox{$7\!\!+\!\!2\!\!+\!\!2(9)\!\!$\rule[-3pt]{0pt}{10pt}}\atop\mbox{$+4\!\!+\!\!11\!\!+\!\!6$\rule[-3pt]{0pt}{8pt}}&
\mbox{$2\!\!+\!\!4\!\!+\!\!6\!\!+\!\!7$\rule[-3pt]{0pt}{10pt}}\atop\mbox{$\!+2(9)\!\!+\!\!11$\rule[-3pt]{0pt}{8pt}}&
3\!\!+\!\!8\!\!+\!\!10&2\!\!+\!\!4\!\!+\!\!2(9) \rule{0pt}{6pt}\\
\hline
10&10&9\!\!+\!\!11&8\!\!+\!\!10\!\!+\!\!12&7\!\!+\!\!9&8&9&4\!\!+\!\!9&
%8\!\!+\!\!3\!\!+\!\!10\!\!+\!\!5\!\!+\!\!12
\mbox{$3\!\!+\!\!5\!\!+\!\!8$\rule[-3pt]{0pt}{10pt}}\atop\mbox{$+10\!\!+\!\!12$\rule[-3pt]{0pt}{8pt}}
&
%7\!\!+\!\!2\!\!+\!\!2(9)\!\!+\!\!4\!\!+\!\!11\!\!+\!\!6
%\mbox{$7\!\!+\!\!2\!\!+\!\!2(9)\!\!$\rule[-3pt]{0pt}{10pt}}\atop\mbox{$+4\!\!+\!\!11\!\!+\!\!6$\rule[-3pt]{0pt}{8pt}}
\mbox{$2\!\!+\!\!4\!\!+\!\!6\!\!+\!\!7$\rule[-3pt]{0pt}{10pt}}\atop\mbox{$\!+2(9)\!\!+\!\!11$\rule[-3pt]{0pt}{8pt}}
&
%11\!\!+\!\!8\!\!+\!\!3\!\!+\!\!10\!\!+\!\!12
\mbox{$1\!\!+\!\!3\!\!+\!\!8$\rule[-3pt]{0pt}{10pt}}\atop\mbox{$\!+10\!\!+\!\!12$\rule[-3pt]{0pt}{8pt}}
&2\!\!+\!\!9&3\!\!+\!\!8\!\!+\!\!10 \rule{0pt}{6pt}\\
\hline
11&11&10&9&8&7&12&5\!\!+\!\!12&4\!\!+\!\!9&3\!\!+\!\!8\!\!+\!\!10&2\!\!+\!\!9&1\!\!+\!\!12&6\!\!+\!\!7\!\!+\!\!11 \rule{0pt}{6pt}\\
\hline
12&12&9&8\!\!+\!\!10&9&12&7\!\!+\!\!11&6\!\!+\!\!7\!\!+\!\!11&3\!\!+\!\!8\!\!+\!\!10&2\!\!+\!\!4\!\!+\!\!2(9)&3\!\!+\!\!8\!\!+\!\!10&6\!\!+\!\!7\!\!+\!\!11&1\!\!+\!\!5\!\!+\!\!2(12)\rule{0pt}{6pt}\\
\hline
\end{array}\nonumber$
\end{center}
\caption{Cayley table of the (commutative) $E_6$ Ocneanu algebra of seams with the compact notation $\eta=\widehat{\mathbf P}_{\eta}$ with $\eta=1,2,\ldots,12$. 
The fusion of these seams reproduces the known $E_6$ Ocneanu Cayley table~\cite{CoqHuerta2003}. 
The top-left quadrant is the Cayley table of the $E_6$ graph fusion algebra.\label{E6Cayley}}
\end{figure}

\subsection{$E_6$ Ocneanu algebra: quotient ring and quantum dimensions}

There is an injective ring homomorphism
\begin{align}
 \phi:\big\langle \widehat{\mathbf P}_{a,1},\widehat{\mathbf P}_{a,2}\,|\,a=1,\ldots,6\big\rangle\to\mathbb{Z}[x,y]\big/\big\langle p_1(x),p_2(x,y)\big\rangle
 \label{quotientring}
\end{align}
where 
\begin{align}
 p_1(x)&=\prod_{\ell\in\Exp(E_6)}\!\!\big(x-2\cos\tfrac{\ell\pi}{g}\big)=x^6-5x^4+5x^2-1\nonumber\\
 p_2(x,y)&=y^2+(x^5-5x^3+4x)y-1,\qquad p_2(y,y)=p_1(y)\label{p2xy}
\end{align}
and
\begin{subequations}
\label{seamPolys}
\begin{align}
 &\widehat{\mathbf P}_{1,1}\mapsto1,\quad
 \widehat{\mathbf P}_{2,1}\mapsto x,\quad
 \widehat{\mathbf P}_{3,1}\mapsto x^2-1,\quad
 \widehat{\mathbf P}_{4,1}\mapsto x^5-4x^3+2x
\\[.15cm]
 &\widehat{\mathbf P}_{5,1}\mapsto x^4-4x^2+2,\quad
 \widehat{\mathbf P}_{6,1}\mapsto -x^5+5x^3-4x,\quad
 \widehat{\mathbf P}_{a,2}\mapsto \phi(\widehat{\mathbf P}_{a,1})\,y,\quad a=1,2,\ldots,6\label{seamPolys2}
\end{align}
\end{subequations}

The $E_6$ integrable seams all mutually commute and are simultaneously diagonalizable. The eigenvalues therefore yield a 1-dimensional representation of the $E_6$ Ocneanu algebra in terms of $s\ell(2)$ quantum dimensions on $\mbox{Vir}\otimes\overline{\mbox{Vir}}$. 
For the spin-$\tfrac 12$ fundamentals, let us set $x=q^s+q^{-s}$, $y=q^{\bar s}+q^{-\bar s}$ with $q=e^{\pi i/12}$ to ensure the fusion rules truncate correctly. 
We then need to solve
\be
p_1(x)=0,\qquad p_2(x,y)=0
\ee
These equations admit $6\times 2=12$ solutions for $x=q^s+q^{-s}=2\cos \tfrac{s\pi}{12}$ and $y=q^{\bar s}+q^{-\bar s}=2\cos \tfrac{\bar s\pi}{12}$ 
where,  explicitly,
\be
(s,\bar s)\in{\cal S}=\{(1,1),(4,4),(5,5),(7,7),(8,8),(11,11),(1,7),(7,1),(4,8),(8,4),(5,11),(11,5)\}\label{ssSectors}
\ee
These indices coincide with the operator content of the $E_6$ modular invariant partition function $Z(q)=\sum_{r=1}^{g-2}\sum_{(s,\bar s)\in{\cal S}} \chi_{r,s}(q)\chi_{r,\bar s}(\bar q)$ for which the conformal weights satisfy $\Delta_{r,s}\!-\!\Delta_{r,\bar s}\in{\mathbb Z}$. As required, the set $\cal S$ is symmetric under interchange of $s$ and $\bar s$. 
In the $(s,\bar s)$ sector, the corresponding quantum dimensions are
\be
\begin{array}{ll}
(1,1)\!: 1\qquad&(1,2)\!: q^{\bar s}+q^{-\bar s}\\
(2,1)\!: q^s+q^{-s}\qquad&(2,2)\!: q^{s+\bar s}+q^{s-\bar s}+q^{-s+\bar s}+q^{-s-\bar s}\\
(3,1)\!: q^{2s}+1+q^{-2s}\qquad&(3,2)\!:  q^{2s+\bar s}+q^{2s-\bar s}+q^{\bar s}+q^{-\bar s}+q^{-2s+\bar s}+q^{-2s-\bar s}\\
(4,1)\!: q^{5s}\!+\!q^{3s}\!+\!q^{-3s}\!+\!q^{-5s}\quad&(4,2)\!: q^{5s+\bar s}\!+\!q^{5s-\bar s}\!+\!q^{3s+\bar s}\!+\!q^{3s-\bar s}\!+\!q^{-3s+\bar s}\!+\!q^{-3s-\bar s}\!+\!q^{-5s+\bar s}\!+\!q^{-5s-\bar s}\\
(5,1)\!: q^{4s}+q^{-4s}\qquad&(5,2)\!: q^{4s+\bar s}+q^{4s-\bar s}+q^{-4s+\bar s}+q^{-4s-\bar s}\\
(6,1)\!: -q^{5s}\!+\!q^s\!+\!q^{-s}\!-\!q^{-5s}\quad&(6,2)\!: -q^{5s+\bar s}\!-\!q^{5s-\bar s}\!+\!q^{s+\bar s}\!+\!q^{s-\bar s}\!+\!q^{-s+\bar s}\!+\!q^{-s-\bar s}\!-\!q^{-5s+\bar s}\!-\!q^{-5s-\bar s}
\end{array}\label{ssbar}
\ee
Taking the union over the 12 sectors (\ref{ssSectors}), we thus find that the distinct eigenvalues of the integrable seams are given by the roots of the minimal polynomials
\be
\begin{array}{ll}
(1,1)\!: \ \ \lambda-1\qquad&(1,2)\!: \ \ p_1(\lambda)\\
(2,1)\!: \ \ p_1(\lambda)\qquad&(2,2)\!: \ \ \lambda^4-4\lambda^3+4\lambda-1\\
(3,1)\!:\ \  \lambda^3-2\lambda^2-2\lambda\qquad\quad&(3,2)\!: \ \ \lambda^7-30\lambda^5+60\lambda^3-8\lambda\\
(4,1)\!: \ \ p_1(\lambda)\qquad&(4,2)\!:\ \  \lambda^4-4\lambda^3+4\lambda-1\\
(5,1)\!: \ \ \lambda^2-1\qquad&(5,2)\!: \ \ p_1(\lambda)\\
(6,1)\!: \ \ \lambda^3-2\lambda\qquad&(6,2)\!: \ \ \lambda^3-2\lambda^2-2\lambda
\end{array}
\ee
The toric seam matrices satisfy the polynomial equations obtained by setting the associated minimal polynomial to zero. We observe that some integrable seams have a zero eigenvalue and so are not invertible. Strictly speaking, in this sense, these seams are not symmetries --- they are ``non-invertible symmetries'' \cite{NonInvSymm}.

\subsection{$E_6$ Ocneanu algebra of seams and structure constants}\label{sec:quotient}

For $E_6$, it follows from (\ref{p2xy}) and Cayley-Hamilton that 
\be
p_1({\mathbf B})=p_1(\overline{\mathbf B})={\mathbf 0}
\ee
Additionally, for system sizes $M\le 12$, we have verified in Mathematica that 
\bea
p_2({\mathbf B},\overline{\mathbf B})=\overline{\mathbf B}^2-\widehat{\mathbf P}_{6,1} \overline{\mathbf B}-{\mathbf I}={\mathbf 0}\label{p2B}
\eea
From (\ref{p2xy}) and (\ref{seamPolys2}), $\mathbf B$ only enters this polynomial equation via the combination
\be
\widehat{\mathbf P}_{6,1}=\widehat{\mathbf N}_6=-{\mathbf B}^5+5{\mathbf B}^3-4{\mathbf B}
\ee
so (\ref{p2B}) can be rewritten as
\be
\overline{\widehat{\mathbf N}}_2^{\,2} = {\mathbf I} + {\widehat{\mathbf N}_6} {\overline{\widehat{\mathbf N}}_2}\label{reduction}
\ee
This is in accord with the product $7*7=1+12$ in the Cayley table of Figure~\ref{E6Cayley}.

Let $a,b,c=1,2,\ldots,6$ and $a',b',c'=1,2$. Armed with the reduction (\ref{reduction}), the product of any two seams $\widehat{\mathbf P}_{a,a'}$ and $\widehat{\mathbf P}_{b,b'}$ can now be obtained using commutativity and associativity. We can do this separately for the four quadrants $(a',b')$ of the Cayley table Figure~\ref{E6Cayley} with $a',b'=1,2$. But, since by commutativity the $(1,2)$ and $(2,1)$ quadrants agree, we only need to consider the three cases $a'b'=1,2,4$:
\begin{align}
a'b'=1:\quad& \widehat{\mathbf N}_a \widehat{\mathbf N}_b=\sum_c \widehat{N}_{ab}{}^c\, \widehat{\mathbf N}_c=
\sum_c \widehat{N}_{ab}{}^c\, \widehat{\mathbf P}_{c,1}\\
a'b'=2:\quad&(\widehat{\mathbf N}_a \overline{\widehat{\mathbf N}}_2) \widehat{\mathbf N}_b=\widehat{\mathbf N}_a (\widehat{\mathbf N}_b\overline{\widehat{\mathbf N}}_2)
=\sum_c \widehat{N}_{ab}{}^c (\widehat{\mathbf N}_c \overline{\widehat{\mathbf N}}_2)=\sum_c \widehat{N}_{ab}{}^c\, \widehat{\mathbf P}_{c,2}\\
a'b'=4:\quad&(\widehat{\mathbf N}_a \overline{\widehat{\mathbf N}}_2)(\widehat{\mathbf N}_b\overline{\widehat{\mathbf N}}_2)
=\sum_c \widehat{N}_{ab}{}^c\, \widehat{\mathbf N}_c\,\overline{\widehat{\mathbf N}}_2^{\,2}
=\sum_c \widehat{N}_{ab}{}^c\, \widehat{\mathbf N}_c\, ({\mathbf I} + {\widehat{\mathbf N}_6} {\overline{\widehat{\mathbf N}}_2})\\
&\hspace{-10pt}=\sum_c \widehat{N}_{ab}{}^c\, \widehat{\mathbf N}_c+\sum_{c,d} \widehat{N}_{ab}{}^c\,\widehat{N}_{c6}{}^d 
(\widehat{\mathbf N}_d \overline{\widehat{\mathbf N}}_2)
=\sum_c \widehat{N}_{ab}{}^c\, \widehat{\mathbf P}_{c,1}
+\sum_{c,d} \widehat{N}_{ab}{}^c\,\widehat{N}_{c6}{}^d \widehat{\mathbf P}_{d,2} \nonumber
\end{align}
This observation gives explicitly the structure constants $\widetilde{N}_\eta$ of the $E_6$ Ocneanu algebra, as in (\ref{OcneanuNimreps}), in terms of the graph fusion and double graph fusion structure constants
\bea
\widetilde{N}_{a,a'\!;\,b,b'}{}^{c,c'}=(\delta_{a'\!,b'}\delta_{c'\!,1}+\delta_{a'b'\!,2}\delta_{c'\!,2})\widehat{N}_{ab}{}^c
+\delta_{a'b'\!,4}\delta_{c'\!,2}\sum_{d=1}^6 \widehat{N}_{ab}{}^d\widehat{N}_{d6}{}^c
\eea

We deduce that the integrable seams $\widehat{\mathbf P}_{a,b}$ in (\ref{basis}) satisfy the $E_6$ Ocneanu algebra for arbitrary system sizes $M$ once the seam relation $p_2(\mathbf B, \overline{\mathbf B})=\mathbf 0$ is established. 
Its proof follows from (\ref{ssbar}). For a given eigenvector of $\T(u)$, the eigenvalues of $\mathbf B, \overline{\mathbf B}$ are $2\cos \tfrac{s\pi}{12}$, $2\cos \tfrac{\bar s\pi}{12}$ with $(s,\bar s)\in{\cal S}$ as in (\ref{ssSectors}). But again, it is readily verified that
\bea
p_2(2\cos \tfrac{s\pi}{12}, 2\cos \tfrac{\bar s\pi}{12})= 0, \qquad (s,\bar s)\in{\cal S}
\eea
Essentially, the proof that the integrable seams satisfy the $E_6$ Ocneanu algebra for arbitrary system sizes $M$ follows by simultaneous diagonalization since the seam eigenvalues, in the form of the quantum dimensions, satisfy the  $E_6$ Ocneanu algebra.
\goodbreak

%%%%%%%%%%%%%%%%%%%%%%%%%%%%
%
\section{Conclusion}\label{sec:conclusion}
%
%%%%%%%%%%%%%%%%%%%%%%%%%%%%

In this short article, we have developed further the lattice approach to fusion in critical unitary RSOS lattice models. Using Yang-Baxter integrability, it is now possible to construct column transfer matrices (seams) for topological defects that appear in the associated CFT in the continuum scaling limit. Like CFT topological defects, the lattice seams commute and propagate freely along the lattice. 
Remarkably, the lattice seams satisfy the Verlinde algebra, for $A$ theories, and the graph fusion and Ocneanu algebras for $D$ and $E$ theories. In addition, the CFT quantum symmetry is already exhibited by the finite size integrable seams on the lattice. Importantly, fusion is implemented by simple matrix multiplication of square matrices associated with the integrable seams. Solving the boundary inversion relations for $\T_{\!h}(u)$ as in \cite{ChuiEtAl2003}, it is seen that the boundary free energies of the $s,a$ and $b$-type seams vanish whereas this is not the case for the general $r$-type seams. It would be of interest to find the boundary entropies of the seams as well as to find projectors onto the $(s,\bar s)$ and $(a,b)$ sectors of $\T(u)$.

Here we only considered type I theories and focussed on the $A$ series and the exceptional $E_6$ theory of the critical unitary RSOS models on the torus.
Crucially, we have identified the $E_6$ Ocneanu algebra as a quotient polynomial ring in two indeterminates~(\ref{quotientring}). 
We will publish details for all the critical \ade\/ \!RSOS models elsewhere. 

Our new methods provide powerful tools to investigate the Ocneanu algebra of seams for other critical lattice models including models of higher spin or higher rank as well as models defined on the cylinder. Indeed, it would be of interest to apply the new methods to critical nonunitary \ade RSOS models~\cite{FB1985}, fused \ade RSOS models~\cite{DJKMO} such as the superconformal series~\cite{GKO1986,RP2002}, as well as the critical dilute \ade RSOS lattice models~\cite{WNS1992,R1992,WPSN1994} including the critical $(A,G)$ series in branch~2 and tricritical $(G,A)$ series in branch~1~\cite{OBrienP1995}.

%%%%%%%%%%%%%%
\subsection*{Acknowledgments}
%%%%%%%%%%%%%%

The Australian Research Council (ARC) is acknowledged for support under Discovery Project DP200102316. 
We thank Jean-Bernard Zuber for comments and Robert Coquereaux for detailed comments on the manuscript. 
This short article is intended for publication in MATRIX Annals. Work on this project was carried out as part of the MATRIX workshop Mathematics and Physics of Integrability (MPI2024):\\
{\color{blue}https:/\!/www.matrix-inst.org.au/events/mathematics-and-physics-of-integrability}\\

%\appendix

%%%%%%%%%%%%%%%%%%%%%%%%%%%%%%%%%%%%%%%%%
%

\end{document}